\documentclass[prb,aps,twocolumn]{revtex4}

\usepackage{graphicx}
\usepackage{amssymb,amsmath}

\newcommand{\figwidth}{.9\columnwidth}

\begin{document}
\title{Adiabatic-antiadiabatic crossover in a
  spin-Peierls chain}
\author{R. Citro}
\affiliation{ Dipartimento di Fisica ``E. R. Caianiello'' and
Unit{\`a} I.N.F.M. di Salerno\\ Universit{\`a} degli Studi di Salerno, Via S.
Allende, I-84081 Baronissi (Sa), Italy}
\author{E. Orignac}
\affiliation{Laboratoire de Physique Th\'eorique de l'\'Ecole Normale
  Sup\'erieure CNRS-UMR8549 \\ 24, Rue Lhomond  F-75231 Paris Cedex 05
  France}
\author{T. Giamarchi}
\affiliation{D\'epartement de Physique de la Mati\`ere Condens\'ee,  Universit\'e de Gen\`eve\\ 24, quai Ernest-Ansermet  CH-1211
  Gen\`eve, Switzerland}
\date{\today}
\begin{abstract}
We consider an  XXZ spin-1/2 chain coupled to optical  phonons
with non-zero frequency $\omega_0$. In the adiabatic limit (small
$\omega_0$), the chain is expected to spontaneously dimerize and
open a spin gap, while the phonons become static. In the
antiadiabatic limit (large $\omega_0$), phonons are expected to
give rise to frustration, so that dimerization and formation of
spin-gap are obtained only when the spin-phonon interaction is large
enough. We study this
crossover using bosonization technique. The effective action is
solved both by the Self Consistent Harmonic Approximation (SCHA)
and by Renormalization Group (RG) approach starting from a
bosonized description. The SCHA allows to analyze the low
frequency regime and determine the coupling constant associated
with the spin-Peierls transition. However, it fails to describe
the SU(2) invariant limit. This limit is tackled by the RG. Three
regimes are found. For $\omega_0\ll\Delta_s$, where $\Delta_s$ is
the gap in the static limit $\omega_0\to 0$, the system is in the
adiabatic regime, and the gap remains of order $\Delta_s$. For
$\omega_0>\Delta_s$, the system enters the antiadiabatic regime,
and the gap decreases rapidly as $\omega_0$ increases. Finally,
for $\omega_0>\omega_{BKT}$, where $\omega_{BKT}$
is an increasing function of the spin phonon coupling, the spin gap
vanishes via a Berezinskii-Kosterlitz-Thouless transition. Our
results are discussed in relation with numerical and experimental
studies of spin-Peierls systems.
\end{abstract}
\maketitle

\section{Introduction}\label{sec:introduction}

The properties of the spin-Peierls (SP)
state in
quasi-one-dimensional materials has attracted considerable
attention over the last decades since its discovery in the organic
compounds of the TTF and TCNQ
series\cite{review_chain,liu_ttf,ducasse_ttf}, and more recently
in the inorganic compound CuGeO$_3$\cite{hase_cugeo3,boucher_cugeo3_review}. In analogy
to the Peierls instability in quasi-one-dimensional
metals\cite{peierls_inst}, a spin-chain undergoes a SP transition
by dimerizing into an alternating pattern of weak and strong bonds~\cite{pincus_spin-peierls,pytte_sp,cross_spinpeierls}
with the magnetic energy gain compensating the energy loss from the lattice
deformation.

Although this physical picture gives a good qualitative
understanding of the SP phenomenon, the real SP transition is in
fact much more complicated to describe. In particular, the above
picture of SP transition is only valid in the adiabatic regime in
which the frequency of the phonons is negligible compared to the
magnetic energy scales in the system, such as the spin gap or the
exchange interaction $J$. The validity of this approximation
 is clearly dependent on the system
at hand. Recently, it was pointed out that the difference between
CuGeO$_3$ and the other SP compounds consists in the high energy
of the optical phonons involved in the transition, which is of the
order of the exchange integral $J$\cite{braden_structure_cugeo3,braden_no_soft_cugeo3,pouget_spinpeierls_data}). Another feature that
distinguishes CuGeO$_3$  from the other organic SP compounds is
that no softening of the phonon modes  is observed near the
transition. All these findings stem for the fact that an adiabatic
treatment of the phonon subsystem\cite{cross_spinpeierls,pytte_sp}
is inadequate to describe the SP transition in CuGeO$_3$ and an
appropriate treatment of phonons in the antiadiabatic regime is
required\cite{uhrig_sp,gros98_no_soft_rpa}.

Unfortunately, not many analytical methods to study the system of
coupled spin and phonons in the full frequency range are available.
The main difficulty relies in the fact that when the phonon
frequency becomes comparable to  the energy gap in the
spin-excitation spectrum , one is entering a quantum regime in which
quantum fluctuations completely impregnate the ground state. This is
why many of the known studies involving dynamical phonons rely on
numerical methods such as exact diagonalization
(ED)\cite{wellein_ed_sp,buchner_ed_sp,augier98_dynamical_sp,augier_dynamical_sP},
strong coupling expansions\cite{trebst01_speierls_dynamics},
density matrix renormalization group
(DMRG)\cite{caron_dmrg_xy_sp,bursill_mit,white_mit,bursill_dmrg_sp}
or Quantum Monte Carlo (QMC)
simulations\cite{sandvik_mc_sp,mckenzie_mc_sp,kuehne99_spin_phonon,raas02_spinpeierls,onishi_spinp_1d}.
From the analytical point of view various approaches have been
developed, but they work well either in adiabatic or in the
antiadiabatic regime. In the former case, most approaches are based
on the mean-field
approximation\cite{cross_spinpeierls,uhrig_dimerized,tsvelik92_dimerized}.
In the latter case various perturbation studies were performed to
derive an effective spin Hamiltonian\cite{fukuyama_pt,weibe_pt}.
Another approach was developed, based on integrating out the phonon
modes\cite{frad_ep} in order to map the model onto the Gross-Neveu
model\cite{gross_neveu} for which various exact results are
available\cite{zamolodchikov79_smatrices,karowski81_gross_neveu}.
Other approaches are based on the flow-equation
method\cite{uhrig_sp} that works well in the antiadiabatic regime,
or on the unitary transformation method for the XY spin
chain\cite{zheng_sp}.

Since various compounds are rather close to the border between the
two regimes\cite{pouget_spinpeierls_data}, it would be thus highly
desirable to have a good method to tackle the
adiabatic-antiadiabatic crossover. In this paper we provide such a
method. We combine the renormalization group (RG) method and the
self-consistent-harmonic-approximation (SCHA) to study the
adiabatic-antiadiabatic crossover in a spin-1/2 Heisenberg chain
coupled to dynamical phonons. A previous attempt to use the RG to
study this adiabatic-antiadiabatic crossover has been
published\cite{sun00_peierls}. We will comment on the differences of
the two approaches.

The plan of the paper is as follows: in Section~\ref{sec:model} we
introduce the model of spin chain coupled to dynamical phonons,
and write it in the continuum limit using the bosonization
representation. In Section~\ref{sec:scha} we describe a
variational approach, inspired from the self consistent harmonic
approximation (SCHA), and use it to describe the crossover. In the
adiabatic regime we find an expression for the spin-Peierls gap
consistent with the mean-field treatment of Cross and
Fisher\cite{cross_spinpeierls}. In the antiadiabatic regime the
gap is essentially described by a sine-Gordon model. In
section~\ref{sec:rg} we study the crossover using a
renormalization group (RG) method. The RG is specially well
adapted around the Heisenberg isotropic point. This allows to
extract the phase diagram as a function of the strength of the
electron-phonon coupling and the phonon frequency. In
section~\ref{sec:discuss} we discuss the findings of the two
methods, both relative to each other and in connection with
experiments. Conclusions can be found in
Section~\ref{sec:conclusions}, and some technical details have
been put in the appendices.

\section{Model and continuum limit} \label{sec:model}

As a simple model which describes a SP system in the following we
consider an antiferromagnetic spin-1/2 chain coupled to a set of
Einstein oscillators,
\begin{equation}\label{ham}
  H=H_s+H_p+H_{sp},
\end{equation}
with
\begin{eqnarray}\label{hshp}
 && H_s=J \sum_n \mathbf{S}_n \cdot \mathbf{S}_{n+1}, \\
 && H_p= \sum_n \left[\frac{p_n^2}{2m} + \frac m 2 \omega_0^2
   q_n^2\right],
\end{eqnarray}
\noindent where $\mathbf{S}_n$ are spin 1/2 operators, $J>0$,
$[q_n,p_{n'}]=i\delta_{n,n'}$. The quantity $m\omega_0^2=k_e$ is the
stiffness of the Einstein phonon.
The interaction of spins with phonons can be modeled  by:
\begin{equation}\label{hsp}
  H_{sp}=g\sum_n q_n \mathbf{S}_n \cdot
  \mathbf{S}_{n+1},
\end{equation}
The coupling to optical phonons described by (\ref{hsp}) is
adequate for the CuGeO$_3$ since it would correspond to a side
group effect by Germanium atoms as discussed in
Refs.~\onlinecite{khomskii,werner_couplings_cugeo3}.
Acoustic phonons could of course be
treated in a very similar way, but one would have to replace $q_n$
with $(q_{n+1}-q_n)$. Note that some
authors\cite{uhrig_sp,augier_dynamical_sP} prefer to diagonalize
the phonon Hamiltonian in (\ref{hshp}) using the boson operator
$b=\sqrt{m\omega_0/2} q+ip/\sqrt{2m\omega_0}$, and write the
interaction $\tilde{g} \sum_n (b^\dagger_n +b_n) \mathbf{S}_n
\cdot
  \mathbf{S}_{n+1}$. It is obvious that one has
  $\tilde{g}=g/\sqrt{2m\omega_0}$. This remark will be useful when we
  will compare the results of the different approaches.
The adiabatic limit is $\omega_0 \to 0$, $m\to \infty$ with $k_e$
fixed. In that limit, the phonons become classical, i.e. the
$q_n$'s commute with the Hamiltonian (\ref{ham}), and one can
simply minimize the ground state energy with respect to their
expectation value. In that limit, the results of
Ref.~\onlinecite{cross_spinpeierls} are recovered. The opposite
antiadiabatic limit is $\omega_0 \to \infty$. In that limit, one
can integrate out the phonons, and one is left with the
Hamiltonian of a frustrated spin 1/2 chain \cite{fukuyama_pt}. For
a frustration large
enough\cite{haldane_dimerized,affleck_log_corr,lecheminant_revue_1d} 
i.e. for large
enough spin-phonon coupling, a spontaneous dimerization of the
spins takes place, and the system presents a spin
gap. Our purpose
is to provide a unified treatment of these two limits.

To solve the spin-phonon problem, we use first the well-known
Jordan-Wigner transformation to express the spin-operators in
terms of spinless fermions. Thus the Hamiltonian $H_s$ becomes:
\begin{multline}\label{hsferm}
  H_f = -t \sum_n [c^\dagger_{n+1}c_n + \text{h.c.}] + \\
  V \sum_n (c^\dagger_{n+1}c_{n+1}-\frac12)(c^\dagger_{n}c_{n}-\frac 1 2),
\end{multline}
with $t=J/2$ and $V=J$. The spin-phonon Hamiltonian $H_{sp}$ is
transformed into:
\begin{multline}\label{hspferm}
  H_{fp}=g \sum_n q_n\left[ \frac 1 2 (c^\dagger_{n+1}c_n +c^\dagger_n
    c_{n+1})\right. + \\ \left.(c^\dagger_{n+1}c_{n+1}-\frac 1
  2)(c^\dagger_{n}c_{n}-\frac 1 2)\right]
\end{multline}
We now proceed in the standard way to take the continuum limit
(see e.g. Ref.~\onlinecite{giamarchi_book_1d} chap. 6). In the
continuous approximation, (\ref{hspferm}) generates a coupling
between the lattice deformation (phonon mode) and the $q=2k_F=\pi$
component of the charge density, $\rho(2k_F, x)$.

In order to get a continuous description
we separate fast and slow components of the phonon field and
similarly for the fermion fields, so we get the
interaction\cite{note_einstein}
\begin{equation}\label{hfp_cont}
  H_{fp}=i \int dx [q(x)\rho(2 k_F, x) -h.c.].
\end{equation}
We now use the boson representation of one dimensional fermion
operators. In this representation the Hamiltonian $H_f$ becomes
\begin{equation}
 H_{f} = \frac1{2\pi} \int dx u K (\pi\Pi(x))^2 + \frac{u}K
 (\nabla\phi(x))^2
\end{equation}
where the field $\phi(x)$ is related to the density of
fermions\cite{giamarchi_book_1d} and
$[\phi(x),\Pi(x')]=i\delta(x-x')$. We have  $u=\frac \pi 2 J
\alpha$, with $\alpha$ the lattice spacing,
 $K=1/2$, $q_n=q(x=n\alpha)$  and we have kept only the
most relevant terms. Changing the parameter $K$ allows one to explore
the more general case of XXZ spin chains with an easy plane
anisotropy\cite{giamarchi_book_1d}.
The long wavelength part of the
fermion density is $\rho_{q \sim 0}(x) = - \frac1\pi
\nabla\phi(x)$ whereas the higher Fourier components
are\cite{lukyanov_ampl_xxz,affleck_ampl_xxz,orignac04_spingap}
\begin{equation}
\label{eq:umklapp}
 \rho_{2 k_F}(x) = \frac{3}{\pi^2 \alpha}\left(\frac \pi
   2\right)^{\frac 1 4} \cos(2 \phi(x)),
\end{equation}
where we have specialized to an isotropic spin chain. The prefactor in
Eq.~(\ref{eq:umklapp}) has been shown in Ref.~\cite{orignac04_spingap}
to yield a good agreement of the gap calculated within bosonization
and numerical
calculations\cite{papenbrock_dimerized}.
The
Matsubara action for the phonon field has a standard quadratic
form:
\begin{equation}\label{sph}
  S_p= \frac \rho 2 \int dx \int_0^\beta d\tau \lbrack
   (\partial_\tau q)^2 +\omega_0^2 q^2\rbrack,
\end{equation}
where $\rho=m/\alpha$ is the  mass density of the optical phonon
mode, and $q(x=n\alpha)=q_n$ is the lattice deformation field.  In
the first approximation (\ref{hshp}) we have neglected the fact
that the phonon disperses. It can be shown that the dispersion
along the chain leads to insignificant corrections. On the
contrary if the phonons are three dimensional, i.e. if $\omega$
disperses with the transverse momentum then significant changes
can occur. Indeed in that case, since the phonon are three
dimensional they couple the different spin chains and can induce a
three dimensional transition at low temperatures. We will come
back to the case of three dimensional phonons in
Section~\ref{sec:discuss}.

Since the total action is quadratic in the phonon fields, we can
integrate them out to obtain the following bosonized action with a
retarded interaction between the electronic densities:
\begin{widetext}
\begin{equation} \label{eq:finite-freq}
  S=\int dx \int_0^\beta \frac{d\tau}{2\pi K} \left[ u (\partial_x \phi)^2 + \frac
    1 u (\partial_\tau \phi)^2\right] -\frac{g^2}{2 (\pi \alpha)^2 \rho \omega_0^2} \int
  dx \int_0^\beta d\tau \int_0^\beta d\tau' \cos 2 \phi(x,\tau) {\cal D}_{\omega_0,\beta}(\tau-\tau') \cos 2
  \phi(x,\tau'),
\end{equation}
\end{widetext}
where ${\cal D}_{\omega_0,\beta}(\tau)$ is the propagator of an Einstein phonon of
frequency $\omega_0$ corresponding to the action (\ref{sph})
\begin{multline}\label{eq:phonon-propagator}
  {\cal D}_{\omega_0,\beta}(\tau-\tau')=\langle q(\tau)q(\tau')\rangle
  = \\
  \frac {\omega_0}{2} \left[ e^{-\omega_0 |\tau-\tau'|}
    + \frac{2\cosh (\omega_0 (\tau-\tau'))}{e^{\beta \omega_0}-1}\right].
\end{multline}
The action (\ref{eq:finite-freq}) fully describes a one
dimensional spin chain coupled to phonons, and does not rely on
adiabatic or antiadiabatic limit. However, one has to note that
because of the cutoff,  the action (\ref{eq:finite-freq}) is valid
only for $\omega_0\ll u/\alpha$. For higher value of the phonon
frequency $\omega_0$, the phonon propagator
(\ref{eq:phonon-propagator}) must be replaced with
$\delta(\tau-\tau')$. In that case the action
(\ref{eq:finite-freq}) is simply the continuum action of a
frustrated spin chain\cite{haldane_dimerized}, in agreement with
the canonical transformation approach\cite{fukuyama_pt,uhrig_sp}.

The action (\ref{eq:finite-freq}) is of course impossible to solve
exactly.  In order to obtain the physical properties of the system
we will analyze it using two different techniques in the next
sections. The first technique is a self consistent approximation.
Such an approximation, will be very useful to define the various
phases of the system as well as the relevant parameters. As any
variational approximation, although it can be very efficient in
describing the various phases it can only describe the transitions
between these phases approximately. So in order to study the
critical points we use a renormalization group method, building on
the knowledge of the relevant parameters extracted from the SCHA.

\section{Self-consistent harmonic approximation} \label{sec:scha}

To study the action (\ref{eq:finite-freq})  we apply first the
Self Consistent Harmonic Approximation (SCHA) or gaussian
variational
method\cite{feynman_statmech,coleman_equivalence,suzumura_sg,giamarchi_columnar_variat}.
The idea is that the action (\ref{eq:finite-freq}) would be
classically minimized by $\phi = 0$. One can thus expect that the
physics will be dominated by small deviations around this minimum
and approximate the action (\ref{eq:finite-freq}) by a quadratic
action. We thus consider as a trial action
\begin{equation} \label{eq:variatac}
 S_0 = \frac{1}{2\beta\Omega} \sum_{q,\omega} G^{-1}(q,\omega_n)
 \phi^*(q,\omega_n) \phi(q,\omega_n)
\end{equation}
We have to find the propagator $G(q,\omega_n)$ so that
(\ref{eq:variatac}) is the best approximation for
(\ref{eq:finite-freq}). For that we define the variational free
energy
\begin{equation}
  \label{eq:free-variational}
  F_{\text{var.}}=F_0 + \langle S-S_0\rangle_0,
\end{equation}
where:
\begin{eqnarray}
  \label{eq:f0}
  F_0 &=& -\frac 1 {\beta L} \ln Z_0=-\frac 1 {\beta }\sum_{q,\omega_n}\ln G(\,\omega_n), \\
  Z_0 &=& \int D\phi e^{-S_0[\phi]},
\end{eqnarray}
and $\langle \ldots \rangle_0$ represents an average with respect
to the action $S_0$.

The second term in the action (\ref{eq:finite-freq}) can be
rewritten as
\begin{widetext}
\begin{equation}\label{eq:equasplit}
 -\frac{g^2}{2 (\pi \alpha)^2 \rho \omega_0^2}
 \int dx \int_0^\beta d\tau \int_0^\beta d\tau' {\cal D}_{\omega_0,\beta}(\tau-\tau')
  [\cos(2 \phi(x,\tau) + 2 \phi(x,\tau')) + \cos(2 \phi(x,\tau) - 2 \phi(x,\tau'))].
\end{equation}
\end{widetext}
Given that the phonon propagator (\ref{eq:phonon-propagator})
decays for large time difference ($\tau-\tau'$), one can see from
(\ref{eq:equasplit}) that the cosine of the sum is roughly
equivalent to $\sim \cos(4 \phi(x,\tau))$ and  can be responsible
for the opening of a gap in the spectrum, while the cosine of the
difference is $\sim (\tau-\tau')^2 (\nabla_\tau\phi(x,\tau))^2$
and thus will modify the quadratic part of the action.

In the following we consider the gapless ($\Delta=0$) and the
gapful ($\Delta\ne 0$) case separately, at zero temperature. The
gapless case is interesting in connection with systems of
electrons at an incommensurate filling interacting with
phonons\cite{voit_phonon_court,voit_phonon,schulz_ep}. In these
systems, the term $\cos [2\phi(x,\tau)+2\phi(x,\tau')]$ does not
appear in (\ref{eq:equasplit}), leading to $\Delta=0$. The
spin-Peierls problem corresponds to the half-filling case for the
fermions.

\subsection{Incommensurate case}\label{sec:ic-case}

Using (\ref{eq:variatac}) the variational free energy is given by:
\begin{widetext}
\begin{multline}\label{eq:varia-xi}
 F_{var}=\frac 1 2\int \frac{dq}{2\pi}\int
  \frac{d\omega}{2\pi}\left\{\lbrack G_0^{-1}(q,\omega)-G^{-1}(q,\omega)\rbrack G(q,\omega)-\ln
  G(q,\omega)\right\} \\
   -\frac{g^2}{2(\pi \alpha)^2 \rho\omega_0^2} \int_{-\infty}^\infty
  d\tau \frac{\omega_0}{2} e^{-\omega_0 |\tau|}
  \langle \cos \left( 2\phi(0,\tau) -\phi(0,0)\right) \rangle,
\end{multline}
\end{widetext}
where $G_0^{-1}(q,\omega)=\frac{1}{\pi K}
(uq^2+\frac{\omega^2}{u})$ and we used
(\ref{eq:phonon-propagator}) for $\beta=\infty$. Introducing the
propagator for the field $\phi$, $G(x,\tau)=\langle
\phi(x,\tau)\phi(0,0)\rangle$, we can rewrite:
\begin{multline} \label{eq:averages-gauss}
   \langle \cos 2\phi(0,\tau) \cos 2\phi(0,0)\rangle = \\
   \frac 1 2 \exp \left\{-4\int \frac{dq}{2\pi}\int
  \frac{d\omega}{2\pi} G(q,\omega) \lbrack 1-\cos(\omega \tau)\rbrack \right\}.
\end{multline}
Using this expression we obtain:
\begin{widetext}
\begin{multline} \label{eq:varia-g}
 F_{var}=\frac 1 2\int \frac{dq}{2\pi}\int
  \frac{d\omega}{2\pi}\lbrack G_0^{-1}(q,\omega)G(q,\omega)-\ln
  G(q,\omega)\rbrack \\
 -\frac{g^2}{4(\pi \alpha)^2 \rho\omega_0^2} \int_{-\infty}^\infty
  d\tau \frac{\omega_0}{2} e^{-\omega_0 |\tau|}
  \exp \left\{-4\int \frac{dq}{2\pi}\int
  \frac{d\omega}{2\pi} G(q,\omega) \lbrack 1-\cos(\omega \tau)\rbrack
     \right\}.
\end{multline}
\end{widetext}
Minimizing the action (\ref{eq:varia-g}) with respect to
$G(q,\omega)$ and using the fact that
$G^{-1}(q,\omega)=G^{-1}_0(q,\omega)-\Sigma (q,\omega)$, where
$\Sigma$ is the self-energy, we get:
\begin{widetext}
\begin{multline} \label{eq:varia-der}
 \frac{\delta F_{var}}{\delta G(q,\omega)}=0=\frac 1 2\int \frac{dq}{2\pi}\int
  \frac{d\omega}{2\pi}\lbrack \Sigma(q,\omega)+  \\
 \frac{g^2}{(\pi \alpha)^2 \rho\omega_0^2} \int_{-\infty}^\infty
  d\tau \frac{\omega_0}{2} e^{-\omega_0 |\tau|} (1-\cos (\omega \tau))
  \exp \left\{-4\int \frac{dq}{2\pi}\int
  \frac{d\omega}{2\pi} G(q,\omega) \lbrack 1-\cos(\omega \tau)\rbrack
     \right\} \rbrack.
\end{multline}
\end{widetext}
As is obvious from the above equation, the low frequency behavior of
$G(q,\omega)$ is similar to the one of $G^{-1}_0(q,\omega)$, and
corresponds to a variational action of the form
\begin{eqnarray}
  \label{eq:var-incom}
  S_0=\int dx \int_0^\beta \frac{d\tau}{2\pi \bar{K}} \left[\bar{u} (\partial_x \phi)^2 + \frac
    {1}{\bar{u}} (\partial_\tau \phi)^2 \right],
\end{eqnarray}
thus $G^{-1}(q,\omega)=\frac{1}{\pi K}
(\overline{u}q^2+\frac{\omega^2}{\overline{u}})$. In the equation
above we thus have:
\begin{multline} \label{eq:intg}
 \int \frac{dq}{2\pi}\int
  \frac{d\omega}{2\pi} G(q,\omega) \lbrack 1-\cos(\omega
  \tau)\rbrack = \\
  \pi \bar{K}\int \frac{dq}{2\pi}\int
  \frac{d\omega}{2\pi} \frac{\lbrack 1-\cos(\omega
  \tau)\rbrack}{(\bar{u}q^2+\frac{\omega^2}{\bar{u}})}=\frac{\bar{K}}{2}
  \ln (\omega_c \tau),
\end{multline}
where $\omega_c=\frac{u}{\alpha}$, is a frequency cutoff. Using
(\ref{eq:intg}) we can write our variational equation as:
\begin{equation}
\Sigma(q,\omega)+\frac{g^2}{(\pi \alpha)^2 \rho \omega_0^2}
\int_{-\infty}^\infty
  d\frac{(\omega_0 \tau)}{2} e^{-\omega_0 |\tau|} \frac{(1-\cos (\omega
  \tau))}{(\omega_c \tau)^{2 \bar{K}}}=0.
\end{equation}
If we confine to an expansion up to order $\omega^2$ in $(1-\cos
(\omega \tau))$, as requested by the analytical behavior of the
Green's function for $\omega \rightarrow 0$,  we obtain:
\begin{equation} \label{eq:sigma_exp}
\begin{split}
 \Sigma(q,\omega) &= -\frac{g^2}{(\pi \alpha)^2 \rho \omega_0^2}
  \int_{-\infty}^\infty
  d\frac{(\omega_0 \tau)}{2} e^{-\omega_0 |\tau|} \frac{\omega^2
  \tau^2}{(\omega_c \tau)^{2 \bar{K}}}  \\
  &= - \frac{g^2}{(\pi \alpha)^2 \rho \omega_0^2} \left(
  \frac{\omega_0}{\omega_c}\right)^{2\bar{K}}
  \frac{\Gamma(3-2K)}{\omega_0^2}\omega^2,
\end{split}
\end{equation}
where $\Gamma$ is the gamma function. As we see the integral is
convergent when $K<3/2$. Going back to the definition of the
self-energy we have:
\begin{equation}
\label{eq:sigma_om}
\Sigma(q,\omega)=(G_0^{-1}-G^{-1})(q,\omega)=\left( \frac{1}{2\pi
uK}-\frac{1}{2\pi \bar{u}\bar{K}}\right)\omega^2.
\end{equation}
Equating (\ref{eq:sigma_exp}) with (\ref{eq:sigma_om}), and using
the fact that $\frac{u}{K}=\frac{\bar{u}}{\bar{K}}$, we obtain the
following  value of the parameter $K$:
\begin{equation}
\label{eq:k_ren} K^2=\bar{K}^2\lbrack 1+\frac{2K g^2}{\pi u \rho
\omega_0^2} \left(\frac{u}{\alpha
\omega_0}\right)^{2-2\bar{K}}\Gamma(3-2K)\rbrack.
\end{equation}
Expanding around $K$, we obtain the renormalized value of
$\bar{K}$:
\begin{equation}\label{eq:k_ren_pert}
\bar{K}^2\simeq K^2\lbrack 1-\frac{2K
g^2}{\pi u \rho \omega_0^2}\left(\frac{u}{\alpha
\omega_0}\right)^{2-2K}\Gamma(3-2K)\rbrack.
\end{equation}
One thus recovers a Luttinger liquid but with a renormalized value
of the Luttinger parameter $K$. (\ref{eq:k_ren_pert}) implies that
$\bar{K}<K$. A similar result can be obtained via the
renormalization group analysis (see the next section). In a RG
analysis the result (\ref{eq:k_ren_pert}) would correspond to
integrating the RG equation for the coupling constant $g$ assuming
that $K$ is not renormalized and then computing the lowest order
correction to $K$ with the renormalized coupling constant. Our
method thus reproduces in a crude way the renormalization of $K$
downwards. As in
Ref.~\onlinecite{voit_phonons_1d,voit_phonon,schulz_ep}, we find
that the tendency of the system to form charge density waves is
increased.

\subsection{Commensurate case}\label{sec:com-case}

In the commensurate case the derivation of the variational free
energy from (\ref{eq:free-variational}-\ref{eq:varia-der}) remains
the same in the commensurate case. Let's rewrite (\ref{eq:varia-g})
in slightly different way:
\begin{widetext}
\begin{multline} \label{eq:varia-g-comm}
 F_{var}=\frac 1 2\int \frac{dq}{2\pi}\int
  \frac{d\omega}{2\pi}\lbrack G_0^{-1}(q,\omega)G(q,\omega)-\ln
  G(q,\omega)\rbrack \\
 -\frac{g^2}{4(\pi \alpha)^2 \rho\omega_0^2}
 \int_{-\infty}^\infty
  d\tau \frac{\omega_0}{2} e^{-\omega_0 |\tau|}
  \left(\frac 1 2\exp
   \left\{-4\int \frac{dq}{2\pi}\int
  \frac{d\omega}{2\pi} G(q,\omega) \lbrack 1-\cos(\omega \tau)\rbrack
     \right\}+ \right. \\
    \left. \frac 1 2\exp\left\{-4\int \frac{dq}{2\pi}\int
  \frac{d\omega}{2\pi} G(q,\omega) \lbrack 1+\cos(\omega \tau)\rbrack
     \right\}\right).
  \end{multline}
Minimizing this action with respect to $G(q,\omega)$ we obtain the
following expression for the self-energy:
\begin{multline}\label{eq:sigma_var_comm}
  \Sigma(q,\omega) = -\frac{g^2}{(\pi \alpha)^2 \rho\omega_0^2} \int_{-\infty}^\infty
  d\tau \frac{\omega_0}{2} e^{-\omega_0 |\tau|} \left((1-\cos (\omega
  \tau))
  \exp
   \left\{-4\int \frac{dq}{2\pi}\int
  \frac{d\omega}{2\pi} G(q,\omega) \lbrack 1-\cos(\omega \tau)\rbrack
     \right\} \right. \\
 \left. +(1+\cos (\omega
  \tau))
  \exp
   \left\{-4\int \frac{dq}{2\pi}\int
  \frac{d\omega}{2\pi} G(q,\omega) \lbrack 1+\cos(\omega \tau)\rbrack
     \right\} \right)
\end{multline}
\end{widetext}
As is obvious from the above equation (\ref{eq:sigma_var_comm}),
$\Sigma(q,\omega)$ is in fact independent of $q$. Moreover,  we
can use the following expansion for the self-energy:
\begin{equation}
\label{eq:sigma_comm} \Sigma(q,\omega)=-\frac{1}{\pi
\bar{u}\bar{K}}(\Delta^2+\gamma \omega^2).
\end{equation}
In Eq. (\ref{eq:sigma_comm}), the variational parameter $\Delta$ stands
  for the gap caused by
the commensurability, and the variational parameter $\gamma$ stands for the
renormalization of the bare Luttinger exponent $K$. Such a restricted
ansatz is justified by the fact that higher powers of $\omega$  in
$\Sigma(\omega)$ are associated with irrelevant operators in the
action, whereas
$\Delta$ and $\gamma$ correspond respectively to a relevant  and a
marginal operator.  Keeping only $\Delta$ amounts to neglect any
renormalization of $K$ by the spin-phonon interaction.

The self-energy (\ref{eq:sigma_comm}) leads to a Green's function $G(q,\omega)$:
\begin{equation} \label{eq:gf_comm}
 G(q,\omega)=\frac{\pi
 \bar{K}}{\bar{u}q^2+\frac{\omega^2}{\bar{u}}+\bar{u} \Delta^2},
\end{equation}
where $\Delta$ is the mass term. The integral of the Green's
function is:
\begin{equation}
\label{eq:intg1} \int \frac{dq}{2\pi}\int
  \frac{d\omega}{2\pi} G(q,\omega) e^{i\omega
  \tau} = \frac{\bar{K}}{2} K_0(\Delta\bar{u}\tau),
\end{equation}
where $K_0$ is the Bessel function. The corresponding variational
action is \begin{eqnarray}
  \label{eq:variational}
  S_0=\int dx \int_0^\beta \frac{d\tau}{2\pi \bar{K}} \left[ \bar{u} (\partial_x \phi)^2 +
  \frac{1}{\bar{u}} (\partial_\tau \phi)^2 + \frac{\bar{u}}{\xi^2} \phi^2
  \right],
\end{eqnarray}
where $u/\xi=\Delta$ is the gap and $u/K=\bar{u}/\bar{K}$ as no term
$(\partial_x\phi)^2$ is generated from (\ref{eq:equasplit}).

Equating the coefficient of (\ref{eq:sigma_comm}) with that coming
from the expansion for small $\omega$ of
(\ref{eq:sigma_var_comm}), we obtain the following two equations:
\begin{widetext}
\begin{eqnarray}
\label{eq:sigma_const} && \frac{\bar{u}\Delta^2}{\pi
\bar{K}}=\frac{2 g^2}{(\pi \alpha)^2
\rho\omega_0^2}\int_{-\infty}^\infty
  d\tau \frac{\omega_0}{2} e^{-\omega_0 |\tau|}\exp
   \left\{-4\int \frac{dq}{2\pi}\int
  \frac{d\omega}{2\pi} G(q,\omega)(1+\cos \omega \tau)
     \right\} \\
&& \frac{\gamma}{\pi \bar{u} \bar{K}} =\frac{2g^2}{(\pi \alpha)^2
\rho\omega_0^2}\int_{-\infty}^\infty
  d\tau \frac{\omega_0}{2} e^{-\omega_0 |\tau|} \frac{\tau^2}{2}
\lbrack e^{-4\int \frac{dq}{2\pi}\int
  \frac{d\omega}{2\pi} G(q,\omega) \lbrack 1-\cos(\omega \tau)\rbrack
     }
     \label{eq:sigma_om2}
  -e^{-4\int \frac{dq}{2\pi}\int
  \frac{d\omega}{2\pi} G(q,\omega) \lbrack 1+\cos(\omega \tau)\rbrack} \rbrack
\end{eqnarray}
\end{widetext}
Using that $\frac{u}{K}=\frac{\bar{u}}{\bar{K}}$, the l.h.s. of
(\ref{eq:sigma_om2}) can also be rewritten as:
\begin{equation}
\frac{\gamma}{\pi \bar{u} \bar{K}}=-\frac{1}{2\pi
uK}+\frac{1}{2\pi \bar{u}\bar{K}}.
\end{equation}
The two self-consistent equations (\ref{eq:sigma_const}) and
(\ref{eq:sigma_om2}) can be solved analytically in the
antiadiabatic limit ($\omega_0 \gg \Delta$). Using
(\ref{eq:intg1}), and after a straightforward but lengthy
calculation we obtain:
\begin{equation}
\label{eq:k_ren_pert_comm} \bar{K}^2\simeq K^2\lbrack 1-\frac{K
g^2}{\pi u \rho \omega_0^2}\left(\frac{u}{\alpha
\omega_0}\right)^{2-2K}\Gamma(3-2K)\rbrack.
\end{equation}
which is the same change of $K$ than in (\ref{eq:k_ren_pert}). The
system also develops a gap given by
\begin{equation}\label{eq:M_ren}
 \Delta=\frac{u}{\alpha}\left[ \frac{K g^2}{\pi u \rho
 \omega_0^2}\left(\frac{u}{\alpha
 \omega_0}\right)^{2K}\Gamma(1+2K)\right]^\frac{1}{2-4\bar{K}}.
\end{equation}
As we can see from (\ref{eq:k_ren_pert_comm}), for $K>1/2$, we can
have $\bar{K}<1/2$, so that (\ref{eq:M_ren}) can still lead to a
gap provided that $g$ is large enough. Combining the two equations
(\ref{eq:k_ren_pert_comm}) and (\ref{eq:M_ren}), we finally have:
\begin{equation}
\bar{K}^2\simeq K^2\lbrack
1-(\Delta\alpha)^{2-4\bar{K}}\frac{\Gamma(3-2K)}{\Gamma(1+2K)}\rbrack.
\end{equation}
The SCHA thus correctly describes the formation of a gap in the
antiadiabatic limit. As for the incommensurate case the SCHA
captures part of the effects of the renormalization of the
parameters. Note that the SCHA, as any variational method is
efficient in capturing the nature of the ordered phases, but in
order to determine the nature of transition one needs the full RG
analysis. Such an analysis will be discussed in Sec.~\ref{sec:rg}

\subsection{Adiabatic-antiadiabatic crossover in the SCHA} \label{sec:crossover}

Using the SCHA we are now in a position to describe the crossover
from adiabatic to antiadiabatic regime. We will assume that we are
far from the point $K=1/2$ and in fact that we have $K<1/2$. For
the spin chain this would correspond in being in the Ising limit.
In that regime, we can neglect the renormalization of $K$ and take
$K=\bar{K}$ in our variational action. The variational free energy
(\ref{eq:varia-g-comm})  can be written:
\begin{widetext}
\begin{equation}
  \label{eq:F-D}
  F= F_0 -\frac{u}{2\pi K \xi^2} G(0,0)
  -\frac{g^2}{4(\pi \alpha)^2 \rho \omega_0^2}
  \left(\frac { e^\gamma \alpha} {2 \xi} \right)^{2K}\int_0^\infty d\tau {\cal
    D}(\tau) \left[e^{-4G(0,\tau)}+e^{4G(0,\tau)}\right]
\end{equation}
\end{widetext}
where $G(0,\tau)$ is given by Eq. (\ref{eq:intg1}). Using the expansion
for the Bessel function we obtain the following approximate
expression:
\begin{equation} \label{eq:approx}
\begin{split}
  G(0,\tau) &= -\frac{K}{2} \ln \left(
    \frac{\sqrt{(u\tau)^2+\alpha^2}}{\xi} \frac {e^\gamma} 2 \right) \quad \text{if}\, u \tau \ll
  2 e^{-\gamma} \xi, \\
  G(0,\tau) &=0 \quad \text{if}\, u \tau \gg 2 e^{-\gamma} \xi,
\end{split}
\end{equation}
where $\gamma$ is the Euler-Mascheroni
constant\cite{abramowitz_math_functions}.  Using the above expression
for the Green's function, we are able to calculate the variational
free energy (\ref{eq:F-D}), and minimizing it with respect to
$\xi$ we obtain the following variational equation:
\begin{widetext}
\begin{eqnarray}
  \label{eq:saddle-point}
  \frac{u}{4\pi K \xi^2} =\frac{ g^2}{(\pi \alpha)^2 \rho \omega_0^2} \left[
\left(\frac {e^\gamma \alpha} {2 \xi} \right)^{2K} e^{-2 e^{-\gamma}
\frac{\omega_0 \xi}{u}} +\left(\frac {e^\gamma \alpha} {2
\xi}\right)^{4K}\left(\frac {u} {\alpha \omega_0} \right)^{2K}
\gamma(1+2K,\frac{\omega_0
  \xi}{u})\right],
\end{eqnarray}
\end{widetext}
where $\gamma(\cdot,\cdot)$ is the incomplete Gamma
function\cite{abramowitz_math_functions}.

Two interesting limits in equation (\ref{eq:saddle-point}) must be
discussed. If $\frac{\omega_0 \xi}{u} \to 0$, one is in the
adiabatic limit, whereas the antiadiabatic one corresponds to
$\frac{\omega_0 \xi}{u} \to \infty$. In the adiabatic limit, one
sees that the term on the right hand side of
(\ref{eq:saddle-point}) reduces to a contribution $\sim
(a/\xi)^{2K}$, the term depending on the incomplete Gamma function
being zero in that limit.  As a result,  the Cross-Fisher
prediction for the gap\cite{cross_spinpeierls},
\begin{equation}
\label{eq:cross-gap}
 \Delta =\frac u \alpha  \left(\frac{g^2} {2k_e}\right)^{\frac 1 {2-2K}}
\end{equation}
is recovered. In the antiadiabatic limit, the exponential term in
(\ref{eq:saddle-point}) disappears, and the incomplete Gamma
function can be replaced by a Gamma function, leading to the
result for the gap we have found in Sec.~\ref{sec:com-case}, in
(\ref{eq:M_ren}). In this limit, the gap can be understood  as
resulting from a $\cos 4\phi$ interaction induced by integrating
out the phonon modes.

\begin{figure}
  \centerline{\includegraphics[width=\figwidth]{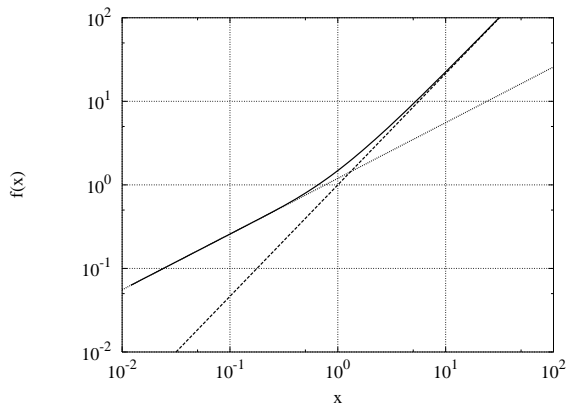}}
\caption{The graph of the function $f(x)$ (solid line) defined in
  Eq. (\ref{eq:definition-f}) with $K=\frac 1 3$.
Two regimes are visible: For $\Delta \gg \omega_0$, $f(x)\sim
x^{2-2K}$ (dashed curve). In that regime, the gap is given by the
adiabatic formula Eq.~(\ref{eq:cross-gap}). For $\Delta \ll
\omega_0$, $f(x)\sim x^{2-4K}$ (dotted
curve) and the gap is given by the antiadiabatic formula
Eq.~(\ref{eq:M_ren}). The crossover regime is observed for
$0.3<\Delta/\omega_0<3$.} \label{fig:graph}
\end{figure}

To perform a general study for any $\omega_0 \xi /u$, we  rewrite
(\ref{eq:saddle-point}) for $\xi$ as:
\begin{widetext}
\begin{eqnarray}
\label{eq:gap-equation1} \frac{\left(\frac \alpha
\xi\right)^{2-2K} }{e^{-2e^{-\gamma}\frac{\omega_0
      \xi}{u}}+\left(\frac {u e^\gamma} {2 \omega_0 \xi} \right)^{2K}
   \gamma(1+2K, 2 e^{-\gamma} \frac{\omega_0 \xi}{u})} =
\frac{4 K g^2}{\pi u \rho\omega_0^2}
\left(\frac{e^\gamma} 2 \right)^{2K}
\end{eqnarray}
In terms of the gap, this equation reads:
\begin{eqnarray}
  \label{eq:gap-equation2}
f\left(\frac{\Delta}{\omega_0}\right) =
\frac{4K g^2}{\pi u
\rho\omega_0^2}\left(\frac{u}{\omega_0
\alpha}\right)^{2-2K}\left(\frac{e^\gamma} 2 \right)^{2K},
\end{eqnarray}
\noindent where:
\begin{eqnarray}
\label{eq:definition-f}
   f(x)=\frac{x^{2-2K}} {e^{-2\frac{e^{-\gamma}}{x}} +\left(x \frac {e^\gamma} {2} \right)^{2K}
  \gamma\left(1+2K,2\,\frac{e^{-\gamma}}{x}\right)}
\end{eqnarray}
\end{widetext}
The graph of the function $f(x)$ is
represented on Fig.~\ref{fig:graph}. In this figure,
the crossover from the adiabatic to the antiadiabatic regime is
easily observed, with the two limiting forms of the gap given
respectively by Eq.~(\ref{eq:cross-gap}) and (\ref{eq:M_ren}).
 The SCHA allows to get the full interpolating
function between the two regimes, and thus to obtain precisely the
crossover scale. We obtain that the
limit between the adiabatic and the antiadiabatic regime is given
by $\omega_0 \sim \Delta $ and not by $\omega_0 \sim J$. This
point will be further discussed in the forthcoming
section~\ref{sec:rg}.

The SCHA also yields the expectation value of the nearest neighbor
correlations ${\bf S}_n\cdot{\bf S}_{n+1}$, as it is proportional
to $(-1)^n \langle \cos 2 \phi\rangle$.  One finds:
\begin{eqnarray}
  \label{eq:expect-dimer}
  \langle S_n \cdot S_{n+1}\rangle \sim \left(\frac \alpha \xi
  \right)^K.
\end{eqnarray}
For $\omega_0\ll \Delta$, i.e. in the adiabatic regime, one has:
\begin{eqnarray}
  \label{eq:adiab-dimer}
  \langle S_n \cdot S_{n+1}\rangle \sim \left(\frac{g^2}{\pi \rho u
      \omega_0^2}\right)^{\frac{K}{2-K}}.
\end{eqnarray}
In the antiadiabatic regime, for $K<1/2$, we find:
\begin{eqnarray}
  \label{eq:antiad-dimer}
  \langle S_n \cdot S_{n+1}\rangle \sim \left[\frac{g^2}{\pi \rho u
      \omega_0^2}\left(\frac{u}{\alpha
        \omega_0}\right)^{2K}\right]^{\frac{K}{2-4K}}.
\end{eqnarray}

\section{RG analysis} \label{sec:rg}

As we have discussed in the previous section, the SCHA  describes only
approximately the renormalization of the quadratic part
by the phonon coupling term. Such a renormalization of the
parameter $K$ is of course specially crucial to take into account
precisely close the isotropic Heisenberg point $K = 1/2$. In this
section, we thus apply an RG method to analyze the
adiabatic-antiadiabatic crossover.

Attempts to an RG analysis of such a problem or of directly related
fermionic problems have been described in the literature. In
particular an RG analysis was performed\cite{augier_dynamical_sP}
at $T=0$ based on a previous work on spinful fermions coupled to
phonons\cite{zimanyi88_eph,caron_eph_2cutoff,bourbonnais96_spinpeierls}.
In this work, the interaction of the spinful fermions with the
electrons is viewed as a retarded backscattering interaction.
However, although this description is appropriate for fermions, in
the case of the spin chain it neglects the fact that the staggered
dimer operator gives rise to more relevant interactions than
current-current ones. As a result, this fermionic description
underestimates the size of the dimerization gap. Our analysis,
directly based on the boson representation of the spin chain does
not suffer from such a limitation. In addition to providing us
with a better description of the dimerization gap,  the use of the
boson representation also allows us to tackle the case of a finite
frequency $\omega_0$ and nonzero temperature.

Another closely related problem is the one of fermions in a random
potential\cite{giamarchi_loc}, which has an action quite similar
to (\ref{eq:finite-freq}) but with a constant ${\cal D}$. One
could be tempted to simply reuse the RG equations derived for this
system. However, here, the situation is more subtle. In
(\ref{eq:phonon-propagator})  for $\omega_0/T \to 0$ , ${\cal
D}(\tau)\to T$. Therefore, we see that the rescaling of the
temperature is going to modify the RG equations with respect to
the case of disordered fermions. Moreover, for $\omega_0/T \to
\infty$, ${\cal D}(\tau)=\omega_0/2 e^{-\omega_0|\tau|}$. As a
result, the limit of $T\to 0$ is delicate to handle properly. In
particular, the definition of the spin-phonon coupling constant
becomes ambiguous in this limit.

However the variational analysis performed in the previous section
allows us to build the correct RG procedure. First the variational
approach shows that in order to obtain the correct results it is
important to first perform the calculation of the ground state
free energy for $0 < T \ll \Delta$, where $\Delta$ is the
spin-Peierls gap, and then take the limit of $T\to 0$. Second it
gives us that the proper dimensionless coupling constant measuring the
strength of the electron-phonon interaction is:
\begin{eqnarray} \label{eq:coupling-constant}
  G=\frac {g^2}{\pi u \rho \omega_0^2}.
\end{eqnarray}
We now proceed with the RG. We start from the following
action:
\begin{widetext}
\begin{multline} \label{eq:action-RG}
  S = \int dx \int_0^\beta \frac{d\tau}{2\pi K} \left[ u (\partial_x \phi)^2+\frac 1 u
    (\partial_\tau \phi)^2\right] - \\
  \frac 1 {2\rho\omega_0^2}
  \left(\frac{g}{\pi a}\right)^2 \int dx \int_0^\beta d\tau
  \int_0^\beta d\tau' \cos 2\phi(x,\tau)
  {\cal D}_{\omega_0,\beta}(\tau-\tau')\cos 2\phi(x,\tau') -\frac{2{g_\perp}}{(2\pi
    a)^2} \int dx \int_0^\beta d\tau \cos 4\phi.
\end{multline}
\end{widetext}
The $\cos(4\phi)$ operator is the marginal operator needed to
describe an spin-isotropic spin chain. The derivation of the
equations is given in Appendix~\ref{app:rg_derivation}. They read:
\begin{subequations}
  \label{eq:RG-spinpeierls}
\begin{eqnarray}
&&  \frac d {dl}\left(\frac 1 K \right) =\left(\frac {g_\perp} {\pi
      u}\right)^2 + \frac{g^2}{\pi u \rho \omega_0^2} \frac \alpha u
  {\cal D}_{\omega_0(l)}(\frac \alpha u), \label{rgK} \\
  \label{eq:RG-spinpeierlsd}
&& \frac d {dl} \left(\frac{{g_\perp}}{\pi u}\right) = (2-4K)
\frac{{g_\perp}}{\pi u} + \frac{g^2}{\pi u \rho \omega_0^2} \frac \alpha
u
  {\cal D}_{\omega_0(l)}(\frac \alpha u), \label{rgdelta} \\
  \label{eq:RG-spinpeierlsg}
&& \frac{d}{dl} \left(\frac{g^2}{\pi u \rho
  \omega_0^2}\right)=\left(2(1-K)+\frac{{g_\perp}}{\pi u}\right)\frac{g^2}{\pi u \rho
  \omega_0^2}, \label{rgGsp} \\
&& \frac {d\omega_0}{dl}=\omega_0. \label{rgomega0}
\end{eqnarray}
\end{subequations}
These RG equations are conveniently expressed using the coupling
constant $G$ defined in Eq.~(\ref{eq:coupling-constant}).
 At this one loop order, we find no corrections to the
phonon frequency as can be seen in (\ref{rgomega0}). However, we
expect such corrections to be obtained in a higher loop order
calculation.

\subsection{Anisotropic case}
Since the action (\ref{eq:finite-freq}) also describes spinless
fermions coupled to phonons, our equations have similarities with
the RG equations that have been derived for the electron phonon
problem\cite{caron_eph_2cutoff,voit_phonons_1d,schulz_ep,voit_phonon,imada_ep}.
There are however important differences. First for a spin chain the
equivalent fermionic band is automatically half-filled (in the
absence of an external magnetic field). Thus in addition to the
standard terms that were considered for the electron-phonon problem
with incommensurate filling, one has here to take into account the
marginal umklapp operator $\cos(4\phi)$ as in
Ref.~\onlinecite{imada_ep}.  Second, in the electron-phonon problem
a different coupling constant is
used\cite{voit_phonons_1d,schulz_ep,voit_phonon}, namely
$Y_{sp}^2=G \frac{\omega_0(l)\alpha} u$. Such definition
appears natural when looking at the RG equations (\ref{rgK}) and
(\ref{rgdelta}) since $Y_{sp}$ seems to be the amount by which $K$
is renormalized in the limit $T=0$. However, such definition would
be at odds with the calculations performed with the SCHA. In fact,
the integral $\int_0^\infty dl \frac \alpha u {\cal D}_{\omega_0
e^l}(\frac \alpha u)=1$  for all $\omega_0$. As a result, if we
neglect ${g_\perp}$ in (\ref{rgK}), and  the renormalization of $K$
in (\ref{rgGsp}), we find the following approximate RG equation for
$G$ and $K$:
\begin{eqnarray}
\label{eq:RG_approx1}
  G(l)&=&G(0) e^{(2-2K)l}, \\
\label{eq:RG_approx2}
  \frac d {dl} (K^{-1})&=& G(0) e^{(2-2K) l} \frac{\omega_0 \alpha}{u}
  e^l \exp\left(-\frac{\omega_0 \alpha}{u} e^l\right),
\end{eqnarray}
and by a variable change to $V=\frac{\omega_0 \alpha}{u}
  e^l$, we easily obtain that
\begin{equation}
\label{eq:renorm-K-approx}
K^{-1}(\infty)-K^{-1}(0)=G(0) \left(\frac {u}{\omega_0
    \alpha}\right)^{2-2K} \Gamma(3-2K).
\end{equation}
This equation is easily understood: $\omega_0$ gives an energy
cutoff that stops the RG flow of $K$ induced by $G$ at an
energy scale of order $\omega_0=u/\alpha e^{-l^*}$.
We note that it is identical to the SCHA result
(\ref{eq:k_ren_pert}). We thus see that at that scale, $K$ is
renormalized by an amount proportional to $G(l^*)$ and not
$G(l) \frac{\omega_0(l)\alpha} u$ as a result of the
exponential factor in (\ref{eq:RG_approx2}).  This confirms that the
right coupling constant in this theory is $G$ and not
$G\omega_0(l)\alpha/u$. In Ref.~\onlinecite{imada_ep}, the same
prescription was used to define the coupling constant whereas in
Ref.~\onlinecite{sun00_peierls}, the incorrect rescaling of
Ref.~\onlinecite{voit_phonon} was used. As a result, we expect the
conclusions of Ref.~\onlinecite{sun00_peierls} to be incorrect in
the adiabatic regime.

Till now, we have assumed that at the scale $l^*=\ln(u/(\alpha
\omega_0))$, the coupling constant $G(l^*)\ll 1$. If this
assumption breaks down, since the coupling constant
$G(l)=e^{(2-2K)l}G(0)$ one finds a gap
\begin{equation}
 \Delta = \frac u \alpha G(0)^{1/(2-2K)}>\omega_0.
\end{equation}
This gap is in agreement with the SCHA result and with the mean
field theory treatment of Cross and
Fisher~\cite{cross_spinpeierls}. For $K<1/2$, in the antiadiabatic
limit $ \omega_0 \gg u/\alpha $, we know from the SCHA that the
phonons can generate a relevant perturbation $\cos 4 \phi$ and
thus induce a gap\cite{frad_ep}. This effect is also captured in
the RG by (\ref{rgdelta}). This can be seen by a two step
renormalization procedure. In the first step, for $l<l^*=\ln \frac
u {\alpha \omega_0}$, a term ${g_\perp}$ is induced by the RG
flow.  This term is found to be of order:
\begin{multline}
  \label{eq:delta-lstar}
  y(l^*)=\frac{{g_\perp}(l^*)}{\pi u}=G(0) \left(\frac u {\alpha
      \omega_0}\right)^{2-2K} \times \\\left[
    \gamma(2K+1,1)-\gamma(2K+1,\frac{\alpha \omega_0}{u})\right].
\end{multline}
Since $\omega_0 \ll u/\alpha$, we can actually neglect
$\gamma(2K+1,\frac{\alpha \omega_0}{u})$ in Eq.~(\ref{eq:delta-lstar}).
For $l>l^*$, ${\cal D}_{\omega_0(l)}(\alpha/u) \to 0$, and we can drop
$G$ from the RG equations. We then have a simple Kosterlitz
Thouless RG flow which leads to a gap of the form:
\begin{eqnarray}
  \label{eq:gap-RG-antiad}
  \Delta=\frac u \alpha \left[ G(0) \left(\frac u {\alpha
      \omega_0}\right)^{2K} \gamma(2K+1,1)\right]^{\frac 1
      {2-4K}}.
\end{eqnarray}
This gap is in agreement with the SCHA prediction in the
antiadiabatic limit (\ref{eq:k_ren_pert_comm}). Therefore, we see
that SCHA and RG methods agree perfectly, far from the isotropic
point, once the proper coupling constant is used in the RG.

Using our RG equations we can now study the SU(2) invariant limit
for which the SCHA cannot be used, due to importance at that point
of the marginally irrelevant operator $\cos(4\phi)$.

\subsection{SU(2) invariant case} \label{sec:su2-isotropic}

In the isotropic limit, we have :
\begin{eqnarray}
  \label{eq:isotropic}
  K=\frac 1 2 \left(1-\frac{{g_\perp}}{2\pi u}\right).
\end{eqnarray}
This ensures that, in the absence of spin-phonon coupling, the
flow will renormalize to the fixed point $K^* = 1/2$ and
$g_\perp^* =0$. It is then easily seen that the equations
(\ref{rgK}-\ref{rgdelta}) reduce to a single equation for
$y=\frac{{g_\perp}}{\pi u}$. This leads to the following RG flow:
\begin{eqnarray}
  \label{eq:simplified1}
&&  \frac {dy}{dl}= y^2 + G(l)\frac{\omega_0 \alpha}{2 u} e^l
  e^{-\frac{\omega_0 \alpha}{u} e^l} \\
     \label{eq:simplified2}
&&  \frac{dG}{dl}=\left(1 +\frac 3 2 y\right) G
\end{eqnarray}
These RG equations allow for the full interpolation between the
adiabatic and antiadiabatic limit.

The simple analysis of the previous section showed that the gap
should behave as $\Delta = \frac u \alpha G(0)$ in the
adiabatic limit. For the isotropic case, using
(\ref{eq:simplified1}-\ref{eq:simplified2}), we obtain logarithmic
corrections to the gap $\Delta =\frac u \alpha G |\ln
G|^{-3/2}$ resulting from the marginal flow of $y(l)$. These
logarithmic corrections (for details see the
Appendix~\ref{app:logar-corr-scal}) are identical to those
obtained by incorporating the logarithmic corrections to the gap
of the dimerized spin 1/2
chain\cite{black_equ,affleck_log_corr,papenbrock_dimerized} into
the Cross-Fisher mean field theory. This confirms that $G$ is
the right coupling constant to study the formation of the spin
Peierls gap in the adiabatic limit. On the other hand, as
discussed in the previous section, in the antiadiabatic limit, it
is the flow of  $y(l)$ that determines whether or not the gap is
formed. To analyze the flow in the antiadiabatic regime, we can
use the approximation $G(l)=G(0)e^{l}$ i.e. we neglect
the logarithmic corrections to the flow of $G$. We have checked
that this approximation leads to a good agreement with the
numerical study of the RG flow using the fourth order Runge-Kutta
algorithm. Using the previous approximation, the RG flow
(\ref{eq:simplified1})-(\ref{eq:simplified2}) can be reduced to a
Ricatti differential equation (cf. Appendix \ref{app:ricatti})
leading to the following dependence of the gap on $G$:
\begin{eqnarray}
\Delta = \omega_0 e^{\gamma-1} \exp\left[-\frac{2\omega_0
\alpha}{u G(0)}\right],
\end{eqnarray}
\noindent for the case of $y(0)=0$. When $y(0)<0$, it is found
that a gap exists only if:
\begin{eqnarray}\label{eq:condition_gap_antiad}
\frac{u G(0)}{2\omega_0\alpha} > \frac{|y(0)|}{1+|y(0)| \ln
\left(\frac{u e^{1-\gamma}}{\alpha\omega_0}\right)}.
\end{eqnarray}
\noindent The physical content of this equation is transparent. At
the scale $l^*$ such that $\omega_0 e^{l^*} = u/\alpha$, $G(l^*)$
is equal to the l.h.s. of the inequality whereas $|y(l^*)|$ is
equal to the r.h.s. of the inequality. The gap can form only if
the renormalized spin-phonon interaction is stronger than the
renormalized marginal coupling at the energy scale $\omega_0$.
This is in agreement with the two step RG
approach\cite{caron_eph_2cutoff} of the preceding section. When
the condition (\ref{eq:condition_gap_antiad}) is satisfied, the
gap behaves as:
\begin{widetext}
\begin{eqnarray}
\label{eq:gap-antiad-marginal}
 \Delta=\omega_0 e^{-(1-\gamma)}
 \exp\left[\frac{-1}{\frac{uG(0)}{2\omega_0\alpha}+\frac{y(0)}{1-y(0)\ln \frac{u e^{1-\gamma}}{\omega_0\alpha}}}\right]
\end{eqnarray}
\end{widetext}
This expression shows that the gap vanishes as
$\exp(-\mathrm{Ct.}/(G(0)-G^c))$ when the spin-phonon coupling
constant goes to the critical value, indicating that the phase
transition between the gapped phase and the gapless phase in the
antiadiabatic regime is in the Berezinskii-Kosterlitz-Thouless
(BKT) universality class. For fixed $G(0)$, the  Eq.~(\ref{eq:gap-antiad-marginal})  also
indicates that there exists
$\omega_{BKT}$, such that for $\omega_0>\omega_{BKT}$
the gap vanishes via a BKT transition. The implicit equation giving
$\omega_{BKT}$ reads:
\begin{eqnarray} \label{eq:omegaBKT}
  \frac{u G(0)}{2\alpha}=\omega_{BKT} \frac{|y(0)|}{1+|y(0)| \ln
\left(\frac{u e^{1-\gamma}}{\alpha\omega_{BKT}}\right)},
\end{eqnarray}
\noindent which shows that $\omega_{BKT}$ is an increasing function of
the spin-phonon coupling constant.

\begin{figure}[htbp]
  \begin{center}
    \includegraphics[width=\figwidth]{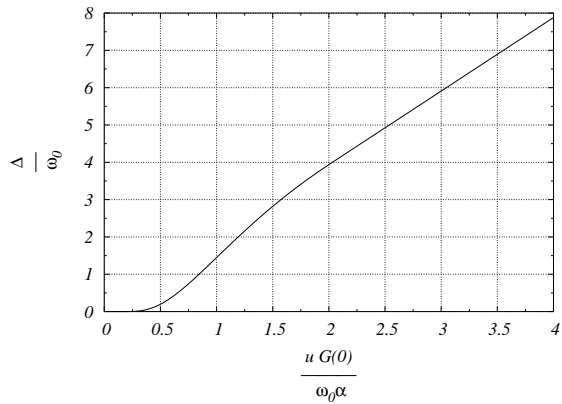}
    \caption{The behavior of the gap $\Delta$ as a function of the spin Peierls
      coupling constant $G$ in the case $y(0)=0$.
      When  $G \gg \alpha \omega_0/u$, the system is in the adiabatic
      regime, and the gap $\Delta$  varies linearly with $G$.
      When $G \ll \alpha \omega_0/u$, the gap decreases very rapidly
      with an essential singularity for $G(0)=0$ described by
      Eq.~(\ref{eq:gap-antiad-marginal}).}
    \label{fig:gap-Gsp}
  \end{center}
\end{figure}

\begin{figure}[htbp]
  \begin{center}
    \includegraphics[width=\figwidth]{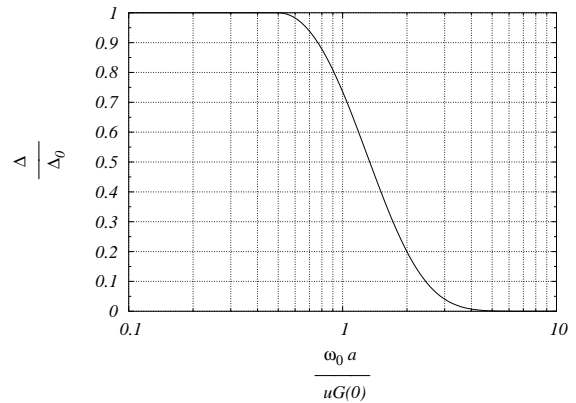}
    \caption{The behavior of the gap $\Delta$ as a function of phonon
      frequency $\omega_0$ for a fixed value of $G(0)$ and
      $y(0)=0$. In
      the adiabatic regime $\omega_0\ll \frac{uG(0)}{\alpha}$,
the gap is independent of $\omega_0$ and equal to
$\Delta_0=\frac{uG(0)}{\alpha}$.
      In the antiadiabatic regime , the gap is a rapidly
      decreasing function of phonon frequency given by
      Eq.~(\ref{eq:gap-antiad-marginal}).}
    \label{fig:gap-w0}
  \end{center}
\end{figure}

\begin{figure}
 \centerline{\includegraphics[width=\figwidth]{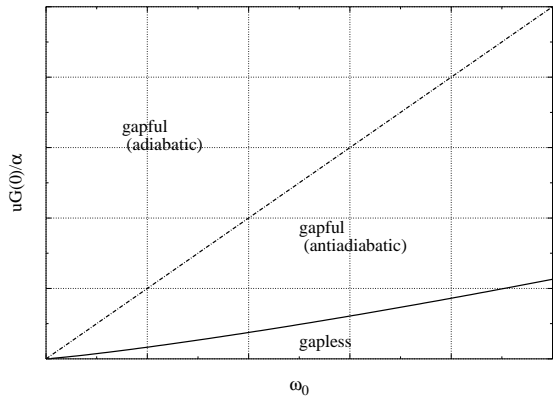}}
 \caption{The phase diagram in the $(G,\omega_0)$ plane, where $G$ is
   the spin-Peierls coupling constant
 $\omega_0$ the phonon frequency.
 The solid line corresponds to the BKT
 transition between the gapped and the
 gapless phase described by Eq.~(\ref{eq:condition_gap_antiad}). The
 dash-dotted line is a crossover line between the antiadiabatic
 regime, $u G(0)/\alpha \ll \omega_0$ and the adiabatic regime $uG(0)
 \gg \omega_0$. } \label{fig:phase_diagram}
\end{figure}

The functional dependence of the gap on the spin-phonon coupling
constant (\ref{eq:gap-antiad-marginal}) is similar to the one
obtained  in the case of a frustrated spin chain. This result is
roughly in agreement with the results of canonical transformations
that eliminate the phonons from the spin
Hamiltonian\cite{fukuyama_pt,uhrig_sp}. Upon closer inspection
however, one finds that the factor of $\frac u {\alpha
  \omega_0}$ does not appear in the formulas giving the spin gap in
that case. The reason is  that the canonical transformations of
Refs.~\onlinecite{fukuyama_pt,uhrig_sp} are valid in the limit
$\omega_0\gg J$ and only the instantaneous interactions are present,
whereas in our theory one needs to renormalize until the scale
$\omega_0$ reaches $J$ before the interactions can be considered
instantaneous . Thus, we find that there is an intermediate regime,
$\Delta <\omega_0 <J$ in which the gap is still larger than in the
frustrated chain limit. The frustrated chain results are recovered
only when $\omega_0 \to J$. In the case of the Heisenberg spin
chain, the value of $y(0)$ has been estimated  to be of the order of
$-0.25$ \cite{affleck_log_corr} which is rather large. For this
reason, we cannot make quantitative predictions on the value of the
gap. However, we can expect from this large value of the marginal
interaction that the critical value of the spin-phonon interaction
needed to obtain the spin gap in the antiadiabatic regime will be
rather large. The behavior of the gap as a function of the strength
of the coupling constant for the case $y(0)=0$
as given by Eq.~(\ref{eq:ansatz-gap-nologs})
is represented on figure~\ref{fig:gap-Gsp}.
The behavior of the gap as a function of frequency is represented on
Fig.~\ref{fig:gap-w0}.
Concerning the expectation value of ${\bf S}_n\cdot {\bf
S}_{n+1}$, we have that $\langle \cos 2 \phi \rangle\sim e^{-K
l_0}$. In the adiabatic regime, we find
\begin{eqnarray}
  \label{eq:dimer-ad-su2}
  \langle \cos 2 \phi \rangle \sim \left(\frac{g^2}{2\pi
      u\rho\omega_0^2}\right)^{1/3},
\end{eqnarray}
whereas, in the antiadiabatic regime, we have:
\begin{multline}
  \label{eq:dimer-anti-su2}
  \langle \cos 2 \phi \rangle \sim \left(\frac{\omega_0
      \alpha}{u}\right)^{1/2} \\
      \exp \left[-\frac{1}{2\left[\frac u
        {\alpha \omega_0} \frac{g^2}{2\pi u \rho \omega_0^2}
        -\frac{|y(0)|}{1+|y(0)|\ln\frac{ue^{1-\gamma}}{\omega_0\alpha}} \right]}\right].
\end{multline}
An ansatz can be made to describe the crossover between the
adiabatic and the antiadiabatic regime.  In the adiabatic regime,
$G(l)\sim 1$ for $l$ such that $\omega_0 e^{l}\ll u/\alpha$ and
$y(l)\ll 1$. We can analyze the crossover from the adiabatic to
the antiadiabatic regime by matching the expressions obtained in
the two cases. This matching procedure (see Appendix
\ref{app:ricatti}) predicts that the crossover of the two regimes
is obtained when $uG(0)/\alpha=2\omega_0=2\omega_c$ (for the case
of $y(0)=0$). The resulting phase diagram is shown in
 Fig.\ref{fig:phase_diagram}, where three regimes are visible.  As
$\omega_0$ increases, we go from the gapped adiabatic regime to
the gapped antiadiabatic regime and finally to the gapless regime.
As is shown the strength of spin-phonon interaction increases the
size of the gapped regime.

\section{Discussion} \label{sec:discuss}

We shall here discuss our results for the spin-Peierls model and
compare with the ones present in the literature. The main purpose
is to give an overview of the informations that can be extracted
using the SCHA and the RG analysis. We stress that our results are
also applicable to strongly interacting fermionic systems
via the Jordan-Wigner
transformation. Moreover, for fermionic systems, both the case of
on-site (Holstein) phonons and on-bond phonons can be dealt with
 as they correspond respectively to a term $\cos 2\phi$ or $\sin
 2\phi$ in the action.

\subsection{Comparison with numerical calculations}

The first result, obtained within SCHA, is the
criterion for the crossover from
the adiabatic to the antiadiabatic regime, which is $\omega_0\sim
\Delta_s$, where $\Delta_s$ is the static gap, and not $\omega_0
\sim J$, as could have been naively been expected. This criterion
was obtained previously by a two-cutoff renormalization group analysis
in Ref.~\onlinecite{caron_eph_2cutoff} in the case of interacting
fermions. Such result is in agreement with a DMRG study of the
XY spin chain coupled to dispersionless phonons of frequency
$\omega_0$, where the crossover is observed when
$\Delta_s\sim \omega_0$.\cite{caron_dmrg_sp} Another
DMRG study in Ref.~\onlinecite{bursill_mit} also showed that in the
case of spinless fermions interacting with Holstein phonons, the
phase transition between the gapful Peierls state (adiabatic) to
the gapless Luttinger liquid state (antiadiabatic) was also obtained when
$\omega_0\sim \Delta_s$.

The second result, obtained using the RG for the SU(2) invariant case
concerns the behavior of the spin gap in the adiabatic and
antiadiabatic cases. We have found that for low frequency,
$\omega_0\ll \Delta_s$, one is in the adiabatic regime with a spin
gap given by the mean-field approximation\cite{cross_spinpeierls}, $\Delta \sim
\Delta_s\sim \frac{g^2}{m\omega_0^2}$, but for $\omega_0\gg \Delta_s$,
in the antiadiabatic regime, the spin gap starts to decrease
rapidly with the frequency. This result is in agreement with the DMRG
study of Ref.~\onlinecite{caron_dmrg_sp}. In fact, in this regime the RG
analysis yields:
\begin{equation}\label{gap-threshold}
 \Delta=\omega_0 e^{-(1-\gamma)}
 \exp\left[\frac{-1}{\frac{uG(0)}{2\omega_0\alpha}+\frac{y(0)}{1-y(0)\ln \frac{u e^{1-\gamma}}{\omega_0\alpha}}}\right]
\end{equation}
\noindent for $\frac {g^2}{2\pi^2 m\omega_0^3} >
|y(0)|/(1+|y(0)|\ln(ue^{1-\gamma}/(\omega_0\alpha)))$, and otherwise the
gap vanishes. In (\ref{gap-threshold}), we have assumed that $y(0)<0$
i.e. that we are dealing with an unfrustrated spin chain which would
not dimerize spontaneously if $g=0$. The three different regimes are
illustrated on Fig.~\ref{fig:phase_diagram}.
An immediate consequence of
(\ref{gap-threshold}) is that
for a sufficiently high frequency $\omega_0$, or for sufficiently weak
 spin phonon
coupling constant a BKT transition to a gapless state  is obtained.
 This transition is analogous to the one that takes place
in the frustrated Heisenberg chain when
$J_2<0.24J_1$\cite{okamoto_frustrated,chitra95_dmrg_frustrated,eggert96_logs,white_zigzag}.

Our
findings for the behavior of the gap, both in the adiabatic and in
the antiadiabatic regime, are in qualitative agreement with the
Quantum Monte Carlo results of Ref.~\onlinecite{onishi_spinp_1d} on
the one-dimensional $S=1/2$ Heisenberg model. There, it was obtained
that for heavy phonon (i.e. low-frequency),  a static gap was
present, while in the case of a light phonon (i.e. high frequency),
no spin gap was observed at the lowest temperature accessible in the
simulation. It is not obvious whether the absence of a dimerization
gap was because the temperature was still above the zero temperature
gap or because of the true absence of a gap above the ground state.
In either case, these results indicate that the spin gap is very
strongly decreased with respect to the static result when the phonon
frequency is increased.
Refs.~\onlinecite{sandvik_mc_sp,sandvik_dyn_phonons}, using
Stochastic Series Expansion methods also found that for small
spin-phonon coupling and $\omega_0/J=1/4$ no spin gap was obtained,
but that increasing the spin phonon coupling above a critical
$\alpha_c=0.225J$ caused a phase transition from the uniform gapless
phase to the dimerized gapped phase, in agreement with our results.
In Ref.~\onlinecite{kuehne99_spin_phonon}, the existence of
dimerization and spin gap was analyzed by Quantum Monte Carlo
simulations of the Heisenberg model for various spin-phonon
couplings and phonon frequencies. A phase diagram (Fig.~9 of
Ref.~\onlinecite{kuehne99_spin_phonon}) was plotted. In agreement
with (\ref{gap-threshold}), it was shown that the critical
spin-phonon coupling to induce dimerization was an increasing
function of the phonon frequency. ED studies
\cite{augier98_dynamical_sp,augier_dynamical_sP} also show
unambiguously the existence of a threshold in the spin-phonon
interaction to induce a spin gap when the phonon frequency is
non-zero. A constant spin gap is obtained when $\tilde{g}$ behaves
roughly as $\omega_0^{1/2}$ which seems in agreement with the
predictions of (\ref{gap-threshold}). Finally, a similar qualitative
agreement is found in the one-dimensional Holstein model for
spin-$\frac{1}{2}$ electrons by DMRG study\cite{white_mit}. The
system undergoes a quantum phase transition between the metallic
phase and the Peierls insulating phase at a finite critical value of
the electron-phonon interaction. Concerning the BKT
universality class of the transition predicted by
(\ref{gap-threshold}),  in the DMRG study of
Ref.~\onlinecite{bursill_dmrg_sp}, it was indeed found that as a
function of the coupling constant the quantum phase transition from
the gapless  to the gapped state is a KT transition. A KT transition
was also found in the related fermionic case  in
Refs.~\onlinecite{caron_dmrg_sp,bursill_mit}. Another result of
Ref.~\onlinecite{bursill_dmrg_sp} regards the evaluation, by
finite-size scaling, of Luttinger exponent $K_\rho$ of the spinless
fermions. It was found that approaching the transition at a critical
value of $g_c$, $K_\rho$ has small deviation from $1/2$ from 0.59 to
0.42. Such finding is in agreement with a BKT
transition driven by the operator $\cos 4\phi$ since the value of
the Luttinger exponent at the transition is then $K=1/2$.

\begin{figure}
 \includegraphics[width=\figwidth]{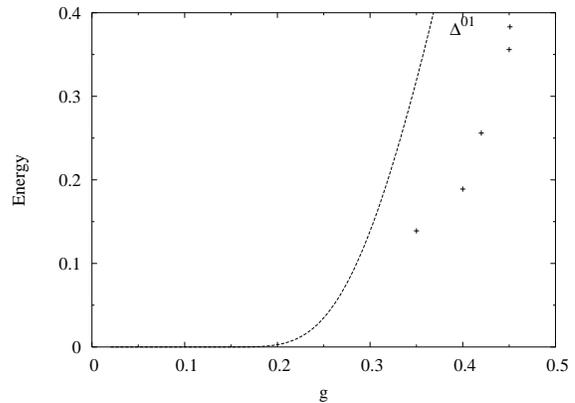}
 \caption{Comparison of our prediction for $\Delta^{01}$ with the
  results of exact
  diagonalization(Ref.~\onlinecite{augier_dynamical_sP}). In both
  cases, the gap is an increasing function of the spin-phonon coupling
  constant. Our results are roughly twice the exact diagonalization
  result.}
  \label{fig:augier-compare}
\end{figure}

In the antiadiabatic regime we predict a power-law relationship
between the critical spin-phonon coupling $g_c$ and the frequency
$\omega_0$. We note that a power law relation between the critical
spin phonon coupling was found in
Ref.~\onlinecite{caron_dmrg_xy_sp}. However, the results of
Ref.~\onlinecite{caron_dmrg_xy_sp} are obtained in the limit of XY
anisotropy, so that a direct comparison of the exponents is not
possible. We can make a more direct comparison with the data of
Ref.~\onlinecite{augier_dynamical_sP}. If we call $g_{DA}$ the spin
phonon  coupling constant used in
Ref.~\onlinecite{augier_dynamical_sP}, it is related to our
spin-phonon coupling constant by:
\begin{eqnarray}
  \label{eq:relation-gs}
  \frac{g^2}{m\omega_0}=4 J^2 g_{DA}^2,
\end{eqnarray}
\noindent yielding for the dimensionless spin-Peierls coupling
constant:
\begin{eqnarray}
  \label{eq:augier-dimensionless}
  G=\frac{4 g_{DA}^2}{\pi^2 \omega_0}.
\end{eqnarray}
The exact diagonalizations in Ref.~\onlinecite{augier_dynamical_sP}
were performed for $\omega_0=0.3J$. The dimensionless parameter
$u/(\alpha\omega_0)=\frac{\pi}{2\times 0.3}\simeq 5.2$, indicating
that a continuum description such as ours should be still
applicable. The values given in
Ref.~\onlinecite{augier_dynamical_sP}, Fig.~1, lead to a
dimensionless coupling constant in the range $[0.17,0.27]$, which is
at the limit of the perturbative regime. Since $\pi
G/2<.429<2\omega_0/J=0.6$, we are in the antiadiabatic regime, not
far from the crossover. The gap we are calculating is $\Delta^{01}$
since our RG approach does not take solitons into account. The
result of the comparison is shown on Fig.~\ref{fig:augier-compare},
where we have replotted the data of
Ref.~\onlinecite{augier_dynamical_sP} for the gap $\Delta^{01}$,
along with the formula (\ref{eq:ansatz-gap-nologs}) in which
logarithmic corrections are neglected.
Obviously, the overall behaviors of the gap with the coupling
constant are very similar. On a more quantitative level there is a
discrepancy between our results and those of
Ref.~\onlinecite{augier_dynamical_sP} by roughly a factor $2$. Such
difference is clearly not due to the logarithmic corrections.
Because our coupling constant is already rather large, the
logarithmic corrections should be rather small. Moreover, if the
discrepancy was caused by logarithmic corrections, it would diminish
as the coupling constant increases, which is not the case here. This
leaves us with three possible explanations of the quantitative
difference between our results and those of
Ref.~\onlinecite{augier_dynamical_sP}. The most likely explanation
is that a coupling constant $G$ in the range $[0.17,0.27]$ is
already a rather large value of the coupling constant, and a
one-loop RG such as ours is not sufficient to obtain the gap
quantitatively in  this regime. A two loops or higher order
calculation may reduce the scale of the gap  with respect to one
loop and lead to better agreement. However, since the usual
techniques for deriving renormalization group
equations\cite{jose_planar_2d,cardy_scaling,knops_sine-gordon} are
restricted to one-loop, one needs to develop a  field theoretic RG
approach\cite{amit_xy}  to check this. Such an approach is beyond
the scope of the present paper. Alternatively numerical data for a
smaller coupling constant would be interesting to compare to. The
second  possible explanation is that our procedure to match the
results of the RG in the adiabatic and the antiadiabatic is
introducing an incorrect scale factor in the antiadiabatic limit.
This is possible if for instance, when we are in still the adiabatic
regime but near the crossover, the interaction with the phonons is
causing a reduction of the gap. Such an effect is ignored in the
Eq.~(\ref{eq:ansatz-gap-nologs}). A possible last explanation is
that because the ratio of the phonon frequency to the exchange
coupling is still not very small, the continuum treatment is not
sufficiently accurate. Note however that despite the relatively
extreme case of this numerical data (large coupling constant and
phonon frequency) with respect to a continuum approach and first
order RG analysis, the quantitative agreement is still reasonably
good.

In addition of providing an analytical framework to describe the
behavior of the spin Peierls gap as a function of coupling constant or
frequency our analysis allows to extract
other physical quantities. In particular a quantity that can be
deduced from our calculations to describe the Peierls ordering
structure of the ground state is the dimerization $\delta= \langle
q_n \rangle$. We can calculate it from the magnetic order
parameter through the relation, $\delta=(-)^n \frac{g}{k_e}\langle
{\mathbf{S}}_i \cdot {\mathbf{S}}_{i+1}\rangle=(-)^n
\frac{g}{k_e}\langle \cos(2\phi) \rangle$. In the adiabatic
regime, the results of the RG analysis gives:
\begin{equation}
\label{eq:dim_ad} \delta \sim (-)^n
\frac{g}{k_e}\left(\frac{g^2}{2\pi
      u\rho\omega_0^2}\right)^{1/3},
\end{equation}
whereas in the antiadiabatic regime we obtain:
\begin{equation}
\label{eq:dim_an} \delta \sim (-)^n
\frac{g}{k_e}\left(\frac{\omega_0
      \alpha}{u}\right)^{1/2} \exp \left[-\frac{1}{2\left[\frac u
        {\alpha \omega_0} \frac{g^2}{2\pi u \rho \omega_0^2}
        -|y(0)|\right]}\right].
\end{equation}
Since by scaling, the expectation value of $\langle \cos(2\phi)
\rangle$ is related to the spin-Peierls gap by the relation $\langle
\cos 2 \phi \rangle\sim (\frac{\Delta}{J})^K$, we immediately obtain
the scaling relation  for the dimerization order parameter, $\delta
\sim \frac{g}{k_e}(\frac{\Delta}{J})^K$. It was found in ED studies
\cite{augier_dynamical_sP} [Fig.1 of
Ref.~\onlinecite{augier_dynamical_sP}] that the alternation $\delta$
in the exchange integral and the spin gap were increasing functions
of the spin phonon coupling constant with a threshold. The
correlation of displacement and dimerization predicted by
(\ref{eq:dim_ad}) and (\ref{eq:dim_an}) are also neatly illustrated
on Fig.~8 of Ref.~\onlinecite{kuehne99_spin_phonon}.

From the behavior of $\langle \cos(2\phi) \rangle$ in the adiabatic
and antiadiabatic regime we can also immediately infer some features
of the spin-spin correlations function. In the antiadiabatic regime,
from (\ref{eq:dimer-anti-su2}), we deduce that for a fixed
frequency, the  correlations decrease at increasing the spin-phonon
coupling [Fig. 15 of  Ref.\onlinecite{kuehne99_spin_phonon}]. At a
fixed spin-phonon coupling and increasing $\omega_0$, the spin-spin
correlation functions become instead less and less affected by $g$.
These findings are again in agreement with the results obtained in
Ref.~\onlinecite{kuehne99_spin_phonon}. We thus see that our
derivation provides a unified framework explaining and generalizing
the previous studies.

\subsection{Relation to experiments}

Let us now turn to experimental systems. Our results clearly show
that non-adiabatic phonon dynamics strongly renormalizes the
magnetic correlations and the dependence of the gap.

\begin{figure}
 \centerline{\includegraphics[width=\figwidth]{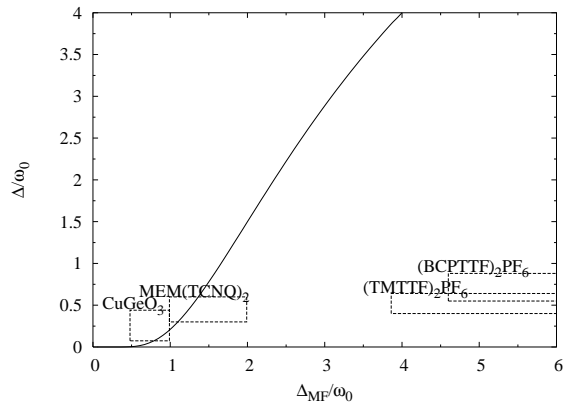}}
 \caption{Various compounds exhibiting a spin-Peierls transition
   plotted on a gap $\Delta/\omega_0$ vs Cross-Fisher gap
   $\Delta_{MF}/\omega_0$.
The data for the gap $\Delta$, the phonon frequency $\omega_0$
 and the Cross-Fisher gap  $\Delta_{MF}$ are extracted from Table~1 of
 Ref.~\onlinecite{pouget_spinpeierls_data}.
 Note that the experimental determination of the gap is to be taken with a grain
 of salt given the differences between the various determinations
 from various measurements (neutrons, thermodynamics, etc.)
 The full line is Eq.~(\ref{eq:ansatz-gap-nologs}) which
   is the dependence of the gap on frequency that we expect when
   logarithmic corrections to scaling can be neglected. }
 \label{fig:comparison_wp}
\end{figure}

A well known example of a material where such strong
renormalizations are observed is the spin-Peierls
material\cite{pouget_spinpeierls_data} CuGeO$_3$. In this material,
the phonon frequency is rather high compared to the actual spin gap
($\omega_0 \sim 310K$). Interestingly, the thermodynamics of this
material can be fitted with a frustrated spin chain
model\cite{castilla_cugeo3}, with $J_2/J_1=0.36$ i.e. well into the
spin gap regime. As was pointed out\cite{uhrig_sp,pouget_spinpeierls_data}
such a dimerization is not intrinsic but due to the spin-phonon
coupling itself. Indeed our RG analysis shows that the low energy
properties of a spin chain coupled to dynamical phonons are similar
at low energy to those of a frustrated spin chain provided that the
phonon frequency is \emph{above} the zero temperature spin gap,
which is the case in CuGeO$_3$. In such a system one can expect a
very strong reduction of the gap due to the finite phonon frequency
as shown in Fig.~\ref{fig:gap-w0}, and was noted before for
CuGeO$_3$. As a point of comparison of the interplay between the
phonon frequency and the phonon coupling constant we have reported
on Fig.~\ref{fig:comparison_wp} the various compounds listed in
Table~1 of Ref.~\onlinecite{pouget_spinpeierls_data}. From
Eq. (\ref{eq:cross-gap}), we expect $\Delta_{MF}\sim \frac u \alpha G(0)$,
and therefore using Eq.~(\ref{eq:ansatz-gap-nologs}), $\Delta=\omega_0
e^{-3.93\frac{\omega_0}{\Delta_{MF}}}$.  Note that the
values of the gap taken here are only indicative, in connection with
our one-dimensional analysis. Indeed, they are (i) dependent on the
measurement method and slightly differ depending which quantity is
measured. (ii) are dependent in part of the interchain couplings
which we have not treated in the present theory.
We nevertheless see on Fig.~\ref{fig:comparison_wp}
that the agreement between our calculated values
of the spin-Peierls gap and the observed one for various systems
both in the antiadiabatic regimes is quite decent.
Although CuGeO$_3$ is the material for which the effects of the
phonon frequency are the stronger, another material for which the
present study could be relevant is MEM(TCNQ)$_2$. For this material
the phonon frequency is of the same order than the spin-Peierls gap
$\Delta \sim 30-60K$, and one can thus still expects effects of the
finite phonon frequency on the spin-Peierls gap. The other two
compounds (TMTTF)$_2$PF$_6$ and (BCPTTF)$_2$PF$_6$ are closer to the
adiabatic regime (since they have $\Delta^{MF}_\sigma > \omega_0$) and
thus are expected
to have a gap less dependent of the phonon frequency than CuGeO$_3$ or
MEM(TCNQ)$_2$. We note that for these two compounds, the agreement
with our formula is not good. For (TMTTF)$_2$PF$_6$, this might result
from the fact that the charge localization temperature  $T\rho\simeq
200$ K is relatively low compared to the spin-Peierls transition
temperature and charge fluctuations can still influence the
transition. Indeed, a description based on adiabatic phonons
interacting with both charge and spin fluctuations can successfully
account for both the magnetic susceptibility and the NMR relaxation
rate in this material\cite{bourbonnais96_spinpeierls}. A related
explanation of the discrepancy could be the existence of a charge
ordering transition\cite{chow_co_1d,monceau_co_1d} in  (TMTTF)$_2$PF$_6$. Such a
transition can affect the mean-field spin-Peierls transition
temperature of the 
material\cite{riera_coexistence_1d,sugiura04_quarterfilled} and thus
invalidate our simple minded estimate of the spin-Peierls coupling
constant. Finally, in both TTF materials, antiferromagnetic
interchain coupling could be relatively important, and they may
diminish the spin-Peierls ordering resulting in a smaller spin-Peierls
gap\cite{inagaki_spinpeierls}.

In order to test for further prediction of the above determination
of the gap it would be interesting to be able to vary continuously
the phonon frequency. Pressure could be an interesting way to
address this question. Since when applying pressure both the
exchange constant and the phonon frequency are to be affected, one
has to compute the net effect on the gap which our theory allows to
do. Such measurements could allow to follow the behavior of the gap
such as described in Fig.~\ref{fig:gap-w0}.

\section{Conclusions} \label{sec:conclusions}

In the present paper, we have analyzed spin-Peierls problem for a
single spin-1/2 chain coupled to an optical phonon of frequency
$\omega_0$ using bosonization techniques. The bosonized action was
approximately solved by using the self-consistent harmonic
approximation.\cite{coleman_equivalence,suzumura_sg} In the low
frequency limit, we have reproduced the result obtained by Cross and
Fisher by the mean-field approximation.\cite{cross_spinpeierls} In
the high frequency limit, we have shown that the retardated
interaction was giving rise to a term that was local in time and was
identical to the bosonized form of a frustrating next-nearest
neighbor interaction in agreement with the canonical transformation
approach.\cite{fukuyama_pt} The self consistent approximation also
allowed us to describe entirely the crossover between the two
regimes. regime that we have considered, the crossover frequency was
given by the spin-Peierls gap $\Delta_s$ calculated for a static
phonon ($\omega\to 0$ with $k_e=m\omega_0^2$ fixed). The adiabatic
regime extends in the region $\omega_0\ll \Delta_s$, and the
antiadiabatic regime extends in the region $\omega_0\gg \Delta_s$.
All the previous findings can be recovered by the renormalization
group by using a two-step approximation as in
Ref.~\onlinecite{caron_eph_2cutoff}. We stress that although we use
the same two-step approximation as
Ref.~\onlinecite{caron_eph_2cutoff} the behavior of the spin gap
that we obtain in the low frequency limit in our renormalization
group is different from the one that one would deduce from the
renormalization group applied to spinful fermions at half-filling
coupled to optical phonons as in
Ref.~\onlinecite{caron_eph_2cutoff}. The reason for this is that in
our problem the charge mode is absent, making the spin-phonon
interaction more relevant than in
Ref.~\onlinecite{caron_eph_2cutoff}. The advantage of the
renormalization group approach over the selfconsistent approximation
is that the former is applicable in the SU(2) invariant case where
the induced non-retarded approximation becomes marginal. For the
SU(2) invariant case, the crossover frequency between the adiabatic
and the antiadiabatic limit remains $\omega_0\sim \Delta_s$.
However, due to the marginality of the induced term, a
BKT transition in the antiadiabatic limit between
the gapped state and the gapless state becomes possible. Near the
transition, the spin gap $\Delta$ drops very rapidly with the phonon
frequency as  $\Delta \sim e^{-C\frac{\omega_0}{\omega_c-\omega_0}}$
and vanishes for $\omega_0>\omega_c$. These results are in
qualitative agreement  with numerical studies\cite{bursill_dmrg_sp}
and with the canonical transformation method.\cite{fukuyama_pt} As
the frequency $\omega_c$ is a increasing function of the spin-phonon
coupling, for fixed $\omega_0$ there exists a critical spin-phonon
coupling below which the spin gap disappears. We have also examined
in the light of the present theory the existing experimental
compounds exhibiting a spin-Peierls transition as summarized for
example in Ref.~\onlinecite{pouget_spinpeierls_data}. We find a good
qualitative agreement with the dependence of the gap predicted by
our theory.

Our analysis thus provides a unified analytical framework in which
to analyze the spin-Peierls transition. It leads the way to
interesting extensions. In particular on could expect to tackle with
similar methods the effects of the interchain couplings, or the
effect of impurities on the spin-Peierls transition (see e.g
Ref.~\onlinecite{pouget-speierls-impurities} and references therein).

\begin{acknowledgments}
E. O. thanks the Physics department of the University of Geneva for
hospitality and support during his visits in May 2003 and October
2003 as well as the Physics department of the University of Salerno
for hospitality in June 2003 and June 2004. R. C. thanks the Physics
department of the University of Geneva for hospitality and support
during her visit in October 2003. Part of this work was supported by
the Swiss National fund for Research under MANEP. E. O. acknowledges
discussions with S. Eggert and M. Mostovoy.
\end{acknowledgments}

\appendix

\section{Expression of the propagator in the SCHA}
\label{app:propagator}

We need to obtain an expression for the propagator $G$. We have:
\begin{eqnarray}
  G(x,\tau) &=& \frac{\pi K u}{\beta} \sum_{\omega_n=2\pi \frac n \beta}
  \int \frac{dq}{2\pi} \frac{e^{i(qx-\omega_n \tau)}}{\omega_n^2 + u^2
      (q^2 +\xi^{-2})} \nonumber \\
      &=& \frac {\pi K}{u} {\cal G}(x,\tau) \label{eq:def-propagator}
\end{eqnarray}
The reduced propagator ${\cal G}$ satisfies to the partial
differential equation:
\begin{eqnarray}
  \label{eq:pde}
  \left[ {u^2} \partial_\tau^2 +\partial_x^2 -\frac 1
    {\xi^2}\right] {\cal G}(x,\tau)=-\delta(x) \sum_{n=-\infty}^\infty
  \delta(\tau-n\beta),
\end{eqnarray}
and to the following properties:
\begin{eqnarray}
  \label{eq:symmetries-G}
  {\cal G}(x,\tau+\beta)&=&{\cal G}(x,\tau) \nonumber \\
  {\cal G}(\pm x,\pm \tau)&=&{\cal G}(x,\tau).
\end{eqnarray}
To solve the partial differential equation (\ref{eq:pde}), we
consider first the auxiliary partial differential equation:
\begin{eqnarray}
  \label{eq:pde-auxiliary}
  \left[ {u^2} \partial_\tau^2 +\partial_x^2 -\frac 1
    {\xi^2}\right] {\cal G}_0(x,\tau)=-\delta(x)\delta(\tau).
\end{eqnarray}
Clearly, if we have a solution of (\ref{eq:pde-auxiliary}), we can
easily deduce from it a solution of (\ref{eq:pde}):
\begin{eqnarray}
  \label{eq:sum}
  {\cal G}(x,\tau)=\sum_{n=-\infty}^\infty {\cal G}_0(x,\tau-n\beta).
\end{eqnarray}
An explicit solution of the PDE (\ref{eq:pde-auxiliary}) is readily
found by Fourier transformation, and application of Eq.~(9.6.21) of
Ref.~\onlinecite{abramowitz_math_functions}. One has:
\begin{eqnarray}
  \label{eq:g0-explicit}
  {\cal G}_0(x,\tau)=\frac u {2\pi}
  K_0\left(\frac{\sqrt{x^2+u^2\tau^2}}{\xi}\right),
\end{eqnarray}
and thus:
\begin{eqnarray}
  \label{eq:g-explicit}
 {\cal  G}(x,\tau)=\frac u {2\pi} \sum_{n=-\infty}^{\infty} K_0 \left(\frac{\sqrt{x^2+u^2(\tau-n\beta)^2}}{\xi}\right).
\end{eqnarray}
It is easily seen that the series in (\ref{eq:g-explicit}) is
convergent, and that the function defined by (\ref{eq:g-explicit})
satisfies all the conditions (\ref{eq:symmetries-G}). The result
(\ref{eq:g-explicit}) could also have been obtained by using the
Fourier transform of the Dirac comb. We note that in reality we
have slightly cheated. Because of the cutoff on the momentum
integral in (\ref{eq:def-propagator}), the function $\delta(x)$ is
in fact smeared out into a function $\delta_\Lambda(x)$ which goes
to the delta function only for $\Lambda \to \infty$. Consequently,
the
 true propagator ${\cal G}_\lambda$ is in fact the convolution of
$\delta_\Lambda(x)$ with the function defined by Eq.
(\ref{eq:g-explicit}). A simple way to incorporate the cutoff in
(\ref{eq:g-explicit}) is to perform the replacement $x^2 \to
x^2+\alpha^2$, with $\alpha \sim \lambda^{-1}$. This substitution
has the advantage of preserving the symmetries
(\ref{eq:symmetries-G}). Finally, we have:
\begin{eqnarray}
  \label{eq:G-final}
  G(x,\tau)=\frac K {2} \sum_{n=-\infty}^{\infty} K_0 \left(\frac{\sqrt{x^2+u^2(\tau-n\beta)^2+\alpha^2}}{\xi}\right).
\end{eqnarray}

\section{Solution of the variational equations} \label{app:varsol}

In the present section, we study the solution of the variational
equations derived from minimization of the free energy
(\ref{eq:F-D}) with respect to $\xi$. The contribution of the region
$\tau \ll \xi/u$ can be written as:
\begin{widetext}
\begin{eqnarray}
  \label{eq:sr-contrib}
&& \frac{g^2}{4(\pi \alpha)^2 \rho \omega_0^2}\left(\frac {e^\gamma \alpha} {2
\xi} \right)^{2K}
 \int_0^{2 e^{-\gamma} \omega_0\xi/u} dv e^{-v} \left[\left(\frac
     {e^\gamma u} {2 \xi
       \omega_0}\right)^{2K} v^{2K}  + \left(\frac {e^\gamma u}  {2 \xi
       \omega_0}\right)^{-2K} v^{-2K}\right] \nonumber \\
&& = \frac{g^2}{4 (\pi \alpha)^2 \rho \omega_0^2}\left(\frac {e^\gamma \alpha}
{2 \xi} \right)^{2K}\left[\left(\frac {e^\gamma u} {2 \xi
      \omega_0}\right)^{2K}\gamma(1+2K,2 e^{-\gamma}
  \frac{\omega_0\xi} u) +
  \left(\frac {e^\gamma u} {2 \xi \omega_0}\right)^{-2K}
  \gamma(1-2K,2 e^{-\gamma} \frac{\omega_0\xi} u)  \right],
\end{eqnarray}
\end{widetext}
where $\gamma$ is the incomplete gamma function (see Ref.
\onlinecite{abramowitz_math_functions} chap. 6 p. 260). Using the
identity:
\begin{eqnarray}
  \label{eq:gamma-identity}
&& \left(\frac \alpha \xi \right)^{2K} \left(\frac u {\xi
    \omega_0}\right)^{-2K}  \gamma(1-2K,2 e^{-\gamma}
\frac{\omega_0\xi}u) \\ && =
\left(\frac {\alpha\omega_0}{u}\right)^{2K}
\left[\Gamma(1-2K)-\Gamma(1-2K,2 e^{-\gamma} \frac{\omega_0\xi}
u)\right]\nonumber
\end{eqnarray}
and noting that the first term in the right hand side is
independent of $\xi$, we can rewrite up to a renormalization the
short-distance contribution to the variational free energy as:
\begin{eqnarray}
  \label{eq:sr-varia}
&&  \left(\frac{u\alpha e^\gamma}{2 \xi^2\omega_0}\right)^{2K}
  \gamma(1+2K,2 e^{-\gamma} \frac{\omega_0 \xi} u)\nonumber \\ &&-\left(\frac{\alpha\omega_0}
    u\right)^{2K}\Gamma(1-2K,2 e^{-\gamma}\frac{\omega_0 \xi}{u})
\end{eqnarray}
The contribution of the region $\tau \gg \frac{\xi}{u}$ can be
rewritten as:
\begin{eqnarray}
  \label{eq:lr-varia}
 && 2\frac{g^2}{4(\pi \alpha)^2 \rho \omega_0^2}\left(\frac {\alpha e^\gamma} {2 \xi} \right)^{2K}
 \int_{2 e^{-\gamma} \omega_0\xi/u}^\infty dv e^{-v}\nonumber \\ && =\frac{g^2}{2(\pi \alpha)^2 m
   \omega_0^2}\left(\frac {\alpha e^\gamma} {2\xi} \right)^{2K} e^{-2 e^{-\gamma}\frac{\omega_0 \xi}{u}}
\end{eqnarray}
Thus, the variational free energy to use reads:
\begin{widetext}
\begin{eqnarray}
  \label{eq:full-varia}
  F&=&F_0 -\frac{u}{2\pi K \xi^2} \langle \phi^2 \rangle -\frac{g^2}{4(\pi
    \alpha)^2 \rho \omega_0^2}\left[2 \left(\frac {\alpha e^\gamma}
      {2\xi}\right)^{2K}e^{-2 e^{-\gamma}\frac{\omega_0 \xi}{u}} +
    \left(\frac {\alpha e^\gamma}
      {2 \xi}\right)^{4K} \left(\frac{u}{\alpha\omega_0}\right)^{2K}
    \gamma(1+2K,2 e^{-\gamma} \frac{\omega_0 \xi}{u})\right. \nonumber
  \\ &&\left. -
    \left(\frac{\alpha\omega_0}{u}\right)^{2K} \Gamma(1-2K,2 e^{-\gamma} \frac{\omega_0
      \xi}{u}) \right]
\end{eqnarray}
\end{widetext}

\section{Derivation of the renormalization group equations}
\label{app:rg_derivation}

To derive the renormalization group equations, we start from the
 action (\ref{eq:action-RG})
where the function ${\cal D}_{\omega_0,\beta}(\tau)$ is defined in
(\ref{eq:phonon-propagator}) and satisfies:
\begin{eqnarray}
  \label{eq:sum_f}
  \int_{-\beta/2}^{\beta/2} {\cal D}_{\omega_0,\beta}(\tau) d\tau =1
\end{eqnarray}
Renormalization group equations are obtained from Operator Product
Expansion techniques (OPE).\cite{cardy_scaling,brunel_random_s=1}
The following OPEs are needed:
\begin{widetext}
\begin{eqnarray}
  \label{eq:opes}
  \cos 2\phi(x,\tau) \cos 2 \phi(x',\tau') &\sim& \frac 1 2 \left[ 1 - \frac
  1 2 \left( 2(x-x') \partial_x \phi(x,\tau) +
    2(\tau-\tau')\partial_\tau \phi(x,\tau)\right)^2 + \cos 4
  \phi(x,\tau)\right] \\
  \cos 2\phi(x,\tau) \cos 4 \phi(x',\tau') &\sim& \frac 1 2 \cos
  2\phi(x,\tau) \\
\cos 4\phi(x,\tau) \cos 4 \phi(x',\tau') &\sim& \frac 1 2 \left[ 1 - \frac
  1 2 \left( 4(x-x') \partial_x \phi(x,\tau) +
    4(\tau-\tau')\partial_\tau \phi(x,\tau)\right)^2\right]
\end{eqnarray}
\end{widetext}
To find the renormalization group equations, we write the partition
function, and expand to second order around the Gaussian fixed
point. The important contributions are of first order in $g^2$ (due to
the nonlocality of the action (\ref{eq:action-RG})), and of
second order in ${g_\perp}^2$ and ${g_\perp} g^2$. Then, we change the
cutoff by $\alpha\to \alpha e^{dl}$. We then apply the OPEs (\ref{eq:opes}) to
obtain the short distance contributions of the terms with
$\alpha^2<(x-x')^2+u^2(\tau-\tau')^2<\alpha^2 e^{2dl}$ to the renormalization of
the coupling constants. Proceeding in that way, we obtain the
following corrections $O(dl)$ to the action:
\begin{widetext}
\begin{eqnarray}
  \label{eq:g2}
  -\frac 1 {2 \rho\omega_0^2}\left(\frac {g}{\pi \alpha}\right)^2
    {\cal D}_{\omega_0(l)}(\frac \alpha
    u)\int_{\alpha<u|\tau-\tau'|<\alpha e^{dl}} dx d\tau d\tau' [-(\tau-\tau')^2
    (\partial_\tau \phi)^2 + \frac 1 2 \cos 4 \phi],
\end{eqnarray}
coming from the $g^2$ term,
\begin{eqnarray}
  \label{eq:delta-delta}
  -\frac 1 {2u} \left(\frac{2{g_\perp}}{(2\pi \alpha)^2}\right)^2
  \int_{\alpha<r <\alpha e^{dl}} r dr
  d\theta \frac 1 2 \left[1-8r^2 \cos^2\theta (\partial_x \phi)^2
    -8 \frac{r^2}{u^2} \sin^2 \theta (\partial_\tau \phi)^2\right],
\end{eqnarray}
coming from the ${g_\perp}^2$ term, and:
\begin{eqnarray}
  \label{eq:delta-g2}
  \frac{-2{g_\perp}}{(2\pi \alpha)^2}\times -\frac{1}{2\rho\omega_0^2}
  \left(\frac {g}{\pi \alpha}\right)^2 \int dx d\tau d\tau' \cos
  2\phi(x,\tau') {\cal D}_{\omega_0(l)}(\tau-\tau')\times  2
  \int_{\alpha<|r-r''|<\alpha e^{dl}}  dx'' d\tau'' \frac 1 2 \cos 2\phi(x,\tau)
\end{eqnarray}
\end{widetext}
coming from the ${g_\perp} g^2$ term. The term (\ref{eq:g2})
contributes to the renormalization of $K,u,{g_\perp}$. The term
(\ref{eq:delta-delta}) contributes to the renormalization of
$u,K$. Finally, the term (\ref{eq:delta-g2}) contributes to the
renormalization of $g^2$. After having performed the mode
integration, we have to restore the original
cutoff\cite{cardy_scaling}. An operator $O$  of scaling dimension
$d_O$ is  rescaled as $O\to Oe^{-d_O l}$ whereas coordinates
are rescaled as $x\to x e^{dl}$ and $\tau \to \tau e^{dl}$. One
also finds $\beta \to \beta e^{-dl}$ i.e. $T\to T e^{dl}$.  As a
result of this operation, we obtain ${g_\perp} \to {g_\perp}
e^{(2-4K) dl}$. The rescaling of $g^2$ is a bit more subtle.
First, we notice that the rescaling of $\alpha \to \alpha e^{dl}$
amounts to a rescaling of $\omega_0,\beta$ inside the function
${\cal D}$. Hence, the rescaling acts on ${\cal D}$ as:
\begin{eqnarray}
  \label{eq:f-rescale}
  {\cal D}_{\omega_0,\beta}\to {\cal D}_{\omega_0(l)e^{dl},\beta e^{-dl}} e^{dl},
\end{eqnarray}
and this way of rescaling ${\cal D}_{\omega_0}$ guarantees that the
constraint (\ref{eq:sum_f}) remains satisfied. The rescaling of
${\cal D}$ absorbs the opposite rescaling of one of the components
$x,\tau,\tau'$. As a result, the rescaling of $g^2$ is given by $g^2
\to e^{(2-2K)dl} g^2$. We notice that this result is in contrast
with the case of a disordered system\cite{giamarchi_loc} in which
the disorder $D$ is rescaled as $D \to D e^{(3-2K)dl}$.
Mathematically, the difference arises because in the case of
phonons, we want to keep the weight in the function ${\cal D}$
constant under the RG flow. This constraint means that physically we
are converting the non-local interaction that exists at high energy
into a local one represented by the $g_\perp$ term by integrating
out successively its short distance contributions. The remaining
weight is then given by $2 \int_{\alpha}^{\beta/2 e^{-l}} {\cal
D}_{\omega e^l,\beta e^{-l}}(\tau) d\tau$ and goes to zero as $l$ is
increased. Adding all the $O(dl)$ contributions, both from
integration of short distance terms and from rescaling, we finally
obtain the renormalization group equations as:
\begin{equation}\label{eq:rges-app}
  \begin{split}
  \frac d {dl} \left(\frac 1 {2\pi u K}\right)&=\frac \alpha u \left( \frac g
    {\pi u}\right)^2 \frac 1 {\rho\omega_0^2} {\cal D}_{\omega_0}(\alpha/u) + \frac{{g_\perp}^2}{2\pi^3
    u^3} \\
 \frac d {dl} \left(\frac u {2\pi  K}\right)&=
 \frac{{g_\perp}^2}{2\pi^3
    u} \\
 \frac{d}{dl} \left(\frac {{g_\perp}}{\pi u}\right)&= (2-4K)\frac
 {{g_\perp}}{\pi u}  + \alpha  \left( \frac g
    {\pi u}\right)^2 \frac 1 {\rho\omega_0^2} {\cal D}_{\omega_0}(\alpha/u)\\
 \frac{d}{dl} \left( \frac{g^2}{\pi u \rho\omega_0^2}
 \right)&=\left(2-2K
  +\frac{{g_\perp}}{\pi u}\right)  \frac{g^2}{\pi u \rho\omega_0^2}\\
 \frac{d\omega_0}{dl}&=\omega_0 \\
 \frac{dT}{dl}&=T
\end{split}
\end{equation}
We note that for $g=0$ these equations reduce to the usual RG
equations of the sine-Gordon model\cite{knops_sine-gordon}. If we
neglect the contribution of ${g_\perp}$ to the renormalization of
$g^2$, we see that the gap obtained for $g^2/(u \rho\omega_0^2) \sim
1$ coincides with the gap predicted by Cross and
Fisher\cite{cross_spinpeierls}. Taking ${g_\perp}$ into account for
$K\sim 1/2$  leads to logarithmic corrections in the dependence of
the gap on $G$. These corrections are discussed in
Appendix~\ref{app:logar-corr-scal}.

\section{Logarithmic corrections to scaling}\label{app:logar-corr-scal}

Let us consider a  spin-1/2 chain with a static dimerization,
described by the Hamiltonian:
\begin{eqnarray}
  \label{eq:dimerized-chain}
  H=J\sum_n (1+\delta(-)^n) \mathbf{S}_n\cdot\mathbf{S}_{n+1},
\end{eqnarray}
Bosonization and scaling arguments
\cite{cross_spinpeierls,haldane_dimerized} lead to the prediction of a
gap $\Delta \sim \delta^{2/3}$. However, the presence of a marginally
irrelevant operator induces corrections to
scaling\cite{black_equ,kadanoff_ashkin,nauenberg_potts,affleck_log_corr,orignac04_spingap}
and the gap behaves in fact as:
\begin{eqnarray}
  \label{eq:gap-dimerized-log}
  \Delta = \frac{1.723 \delta^{2/3}}{\left(1+\frac 2 3 |y(0)| \ln
  \frac {|y(0)|}{1.3612 \delta}\right)}^{1/2}.
\end{eqnarray}
\noindent As a result of logarithmic corrections, the ground state
  energy of the dimerized spin chain behaves as:
  \begin{eqnarray}
    \label{eq:energy-dimerized-log}
     \frac{E_0}{J}=-\frac{0.2728 \delta^{4/3}}{1+ \frac 2 3 |y_2(0)| \ln
    \left|\frac{|y(0)|}{1.3612 \delta}\right|}
  \end{eqnarray}
If we now consider that the chain (\ref{eq:dimerized-chain}) is
coupled to adiabatic phonons, one needs to minimize the total energy $k_e/g^2
\delta^2 - B \delta^{4/3}|\ln \delta|^{-1}$ with respect to $\delta$
yielding  $\delta^{2/3}
|\ln \delta|\sim g^2/k_e$. The gap thus behaves as
\begin{eqnarray}\label{eq:CF-logs}
 \Delta =6.07 \frac{g^2}{m\omega_0^2}  \left|\ln \frac{g^2}{m\omega_0^2}\right|^{-3/2},
\end{eqnarray}
\noindent where we have used the factors quoted in
Eqs.~(\ref{eq:gap-dimerized-log})-~(\ref{eq:energy-dimerized-log}) and
Ref.\cite{orignac04_spingap}.
We know show how the expression (\ref{eq:CF-logs}) can be recovered
within our RG approach.
Using the initial conditions with $SU(2)$ symmetry, the RG equations
are given by the Eqs.~(\ref{eq:simplified1}-\ref{eq:simplified2})

If we now assume that $G(0) \ll y(0)$, in Eq.~(\ref{eq:simplified1}) we
can take $G=0$, so that the previous equations reduce to the
single BKT equation
\begin{equation}\label{kt}
 \frac{dy}{dl}=y^2.
\end{equation}
For $y<0$, this equation flows to a fixed point $y^\star=0$ with
the flow given by:
\begin{equation}\label{flow_y1}
  y(l)=\frac{y(0)}{1-y(0)l}.
\end{equation}
Using (\ref{flow_y1}) in Eq.~(\ref{eq:simplified2}) we can easily
integrate it and obtain:
\begin{equation}\label{flow_y2}
 G(l)=G(0) \frac{e^l}{\lbrack
 1+|y(0)|l \rbrack^{3/2}}.
\end{equation}
This equation should break down for $G(l_0) \simeq |y(l_0)|$ and
after that an exponential flow of $G$ is expected. Using
Eq.~(\ref{flow_y2}), the strong coupling behavior is obtained
when:
\begin{equation}
e^{l^*}=\frac 1 {G(l_0) e^{-l_0}}=\frac{(1+|y(0)|l_0)^{3/2}}{G(0)},
\end{equation}
\noindent where $l_0$ is given by:
\begin{eqnarray}
  \label{eq:l0-def}
  G(0)e^{l_0}=|y(0)|(1+|y(0)|l_0)^{1/2}.
\end{eqnarray}
Solving  Eq.~(\ref{eq:l0-def}) by iteration and using the first
iteration, one  finds the following scaling for the spin-Peierls gap
valid for small $G$:
\begin{equation}\label{eq:RG-gap}
 \Delta_{sp}\simeq e^{-l^\star}=\frac{G(0)}{\left[1+|y(0)| \ln
    \frac{|y(0)|}{G(0)} \right]^{3/2}}.
\end{equation}
This behavior is in agreement with the behavior obtained in
Eq.~(\ref{eq:CF-logs}) by considering logarithmic
corrections to the energy of the dimerized chain.

\section{Study of the SU(2) invariant RG equations}\label{app:ricatti}

We consider the system of 2 coupled first order differential
equations (\ref{eq:simplified1})-(\ref{eq:simplified2}). A
convenient approach in the antiadiabatic regime is to recast this
differential system  as a single differential equation. It is also
convenient to make the approximation $G(l)=G(0)e^l$ as to render
second order equation linear. Introducing\cite{ince_odes}:
\begin{eqnarray}
  \label{eq:definition-Y}
  Y(l)=\exp\left[-\int_0^l y(l') dl'\right].
\end{eqnarray}
We obtain the following second order differential equation:
\begin{eqnarray}
  \label{eq:second-order}
  \frac{d^2Y}{dl^2}&=&-\frac{\omega_0 G(0)\alpha}{2u}  e^{2l}
  e^{-\frac{\omega_0\alpha}{u} e^l} Y(l)\\
&=&-\frac{u G(0)}{2 \omega_0 \alpha} \left( \frac{\omega_0 \alpha
e^l}
  u\right)^2 e^{-\frac{\omega_0\alpha}{u} e^l} Y(l)
\end{eqnarray}
with the initial conditions for $Y(0)=1$ and $Y'(0)=-y(0)$.

It is possible to simplify the second order differential equation
(\ref{eq:second-order}) by a variable change to $s=\frac {\omega_0
\alpha} u e^l$. Writing $Y(l)=Z(\frac {\omega_0 \alpha} u e^l)$,
one finds that $Z(s)$ satisfies to the differential equation:
\begin{equation}\label{eq:Z-equadiff}
  \frac{d^2 Z}{ds^2} + \frac 1 s \frac{dZ}{ds} + \frac
  {uG(0)}{2\omega_0\alpha} e^{-s} Z(s)=0,
\end{equation}
with initial conditions:
\begin{eqnarray}\label{eq:Z-initial}
 Z(\frac{\omega_0\alpha}{u})&=&1, \\
 \frac{\omega_0\alpha}{u}\frac
 {dZ}{ds}(\frac{\omega_0\alpha}{u})&=&-y(0).
\end{eqnarray}
The equation (\ref{eq:Z-equadiff}) can be recast in the form of an
integral equation which reads:
\begin{eqnarray}\label{eq:integ-Z}
 Z(s)&=&1-y(0)\ln\left(\frac{s u}{\omega_0\alpha}\right)\nonumber \\ && -\frac{u
 G(0)}{2\omega_0\alpha} \int_{\frac{\omega_0 \alpha}{u}}^s
 \frac{ds'}{s'} \int_{\frac{\omega_0 \alpha}{u}}^{s''} s'' e^{-s''}
 Z(s'') ds''
\end{eqnarray}
We study first the antiadiabatic limit, $uG(0)/\alpha\ll \omega_0$.
Let us begin with the case $y(0)=0$. By iterating the equation
(\ref{eq:integ-Z}) once, we obtain:
\begin{eqnarray}
  Z(s)&=&1-\frac{u
  G(0)}{2\omega_0}\left[\left(1+\frac{\omega_0\alpha}{u}\right)
  \ln \left(\frac{s u}{\omega_0 \alpha}\right) +
  e^{-s}\right.\nonumber \\
  && \left. -e^{-\omega_0
  \alpha/u} +E_1(s)-E_1\left(\frac{\omega_0 \alpha}u\right)\right]
  \nonumber\\
  & & + o(\frac{u
  G(0)}{2\omega_0}),
\end{eqnarray}
\noindent where $E_1$ is defined in
Ref.~\onlinecite{abramowitz_math_functions}.  Using the fact that
$\omega-0\alpha/u \ll 1$, and $\omega_0\alpha/u e^l \gg 1$, we can
rewrite this equation as:
\begin{equation}
Y(l)=1-\frac{u
  G(0)}{2\omega_0}\left[ l-\ln\left(\frac u
  {\omega_0\alpha}\right) + \gamma -1\right]
\end{equation}
where $\gamma=0.577\ldots$ is Euler-Mascheroni's
constant\cite{abramowitz_math_functions}. The gap is obtained when
$Y(l^*) = 0$, i.e.
\begin{equation}\label{eq:gap-antiad}
  l^*=\ln \left(\frac{u}{\alpha\omega_0}\right) + \frac{2\omega_0\alpha}{u
  G(0)} + 1-\gamma
\end{equation}

The differential equation  (\ref{eq:Z-equadiff}) can also give some
indication on the crossover scale to the adiabatic regime. Assuming
that $\frac{uG(0)}{2\omega_0\alpha}\gg 1$, it is reasonable to
replace the term $e^{-s}$ with $1$ in (\ref{eq:Z-equadiff}),
yielding the approximate equation:
\begin{equation}\label{eq:Z-ed-approx}
 \frac{d^2 Z}{ds^2} + \frac 1 s \frac{dZ}{ds} + \frac
  {uG(0)}{2\omega_0\alpha} Z(s)=0
\end{equation}
The solution of the above differential equation is easily
found\cite{abramowitz_math_functions} in terms of Bessel
functions. One has:
\begin{equation}\label{eq:sol-approx-Z}
  Z(s)=A J_0\left(\left(\frac{u
  G(0)}{2\omega_0\alpha}\right)^{1/2}s\right)+B Y_0\left(\left(\frac{u
  G(0)}{2\omega_0\alpha}\right)^{1/2}s\right)
\end{equation}
The initial conditions yield the following linear system for $A,B$:
\begin{eqnarray}\label{eq:init-sol}
A J_0\left(\left(\frac{\omega_0\alpha
  G(0)}{2u}\right)^{1/2}\right)+B Y_0\left(\left(\frac{\omega_0\alpha
  G(0)}{2u}\right)^{1/2}\right)  &=& 1\nonumber \\
A J_1\left(\left(\frac{\omega_0\alpha
  G(0)}{2u}\right)^{1/2}\right)+B Y_1\left(\left(\frac{\omega_0\alpha
  G(0)}{2u}\right)^{1/2}\right) &=& y(0) \nonumber \\  \times \sqrt{ \frac {2u}{\omega_0
  \alpha G(0)}} &&
\end{eqnarray}
If we consider the case of $y(0)=0$, we find:
\begin{eqnarray}
A=-\frac{\pi}{2} \left( \frac{\omega_0 \alpha G(0)}{2
u}\right)^{1/2} Y_1\left(\left(\frac{\omega_0\alpha
  G(0)}{2u}\right)^{1/2}\right) \\
B=\frac{\pi}{2} \left( \frac{\omega_0 \alpha G(0)}{2
u}\right)^{1/2} J_1\left(\left(\frac{\omega_0\alpha
  G(0)}{2u}\right)^{1/2}\right)
\end{eqnarray}
For $\omega_0 \alpha G(0) /u  \ll 1$, an approximate solution is
$Z(s)=J_0((uG(0)/(2\omega_0\alpha))^{1/2} s)$. We obtain $Z(s)=0$
for $(uG(0)/(2\omega_0\alpha))^{1/2} s=j_{0,1} \simeq 2.40482$ (cf.
Ref.~\onlinecite{abramowitz_math_functions}). The resulting strong
coupling scale would be given by:
\begin{equation}\label{eq:strong-coupling}
  l^*=\ln\left[\frac{j_{0,1}}{G(0)}
  \left(\frac{2u}{\omega_0\alpha}\right) \right]
\end{equation}
We note that at this scale $G(l^*) \gg 1$, but $\frac{\omega_0
\alpha}{u} e^{l^*}\ll 1$. This indicates that $G(l)$ is reaching
strong coupling before the scale $l^*$ is reached, and in this
regime, the gap is produced by $G$ directly. The true strong
coupling scale in this adiabatic regime is then $l^*_{ad}=\ln
(1/G(0))$. The above calculation thus allows to determine when the
crossover from adiabatic to antiadiabatic regime is obtained.

Comparing $l^*_{ad}$ with $l^*_{antiad.}$ we find that their
difference is minimal when $\frac{u G(0)}{\omega_0 \alpha}=2$.
Thus, for $uG(0) \ll \omega_0\alpha$, we are in the antiadiabatic
regime, and:
\begin{equation}\label{eq:gap-antiad1}
\Delta=\omega_0 e^{-(1-\gamma)} e^{-\frac {2\omega_0\alpha}{u
G(0)}}
\end{equation}
For $uG(0)\gg\omega_0\alpha$, the gap behaves as $\Delta= {\cal C} u
G(0)$. We need to match the two results. To do this, we require that
the two expression of the gap are equal and have the same derivative
with respect to $G(0)$ when $uG(0)/\alpha =2\omega_0$. This yields
the following ansatz for the gap:
\begin{eqnarray}
  \label{eq:gap-nolog}
  \Delta = \frac {\cal C} {2e} \frac{u}{\alpha} G(0),
  \frac{u}{\alpha} G(0) \gg \omega_0 \\
  \Delta = {\cal C} \omega_0  e^{- \frac{2\omega_0
      \alpha}{u G(0)}},  \frac{u}{\alpha} G(0) \ll \omega_0
\end{eqnarray}
The constant ${\cal C}$ can be obtained using results for the
adiabatic limit.
Using Eq. (8) of Ref.~\onlinecite{orignac04_tbamf},
we find that ${\cal C}/(2\pi e)=0.627$ leading
to ${\cal C}=10.7$. This leads us to the following ansatz for the gap:
\begin{eqnarray}
\label{eq:ansatz-gap-nologs}
  \Delta=0.627 \frac{g^2}{m\omega_0^2} \; \text{for}\; 2\omega_0 <\frac u \alpha
  G(0) \\
  \Delta=10.7 \omega_0 e^{-2\pi \frac{m\omega_0^3}{g^2}} \;  \text{for}\; 2\omega_0 >\frac u \alpha
  G(0)
\end{eqnarray}
The above ansatz Eq.~(\ref{eq:ansatz-gap-nologs})
tends to overestimate the the gap in the adiabatic regime
as it neglects completely the effect of the non-zero frequency on the
gap. However, the lack of a precise criterion to decide when the RG
flow has reached the strong coupling regime prevents us from finding a
better answer.

We now turn to the case of $y(0)\ne 0$. We first look at the
antiadiabatic limit. We obtain by iterating the equation
(\ref{eq:integ-Z}) that:
\begin{widetext}
\begin{eqnarray}
 Z(s)&=&1-y(0)\ln\left(\frac{us}{\omega_0\alpha}\right)
 -\frac{uG(0)}{2\omega_0\alpha} \left[ \left(1+\frac{\omega_0\alpha}{u}\right)
 e^{-\frac{\omega_0\alpha}{u}}
 -y(0)\left(e^{-\frac{\omega_0\alpha}{u}}+E_1\left(\frac{\omega_0\alpha}{u}\right)\right)\right]\ln
 \left(\frac{us}{\omega_0\alpha}\right) \nonumber\\
 && +\frac{uG(0)}{2\omega_0\alpha}
 \left[1-y(0)\ln\left(\frac{u}{\omega_0\alpha}\right)\right]
 \left(E_1\left(\frac{\omega_0\alpha}{u}\right)+e^{-\frac{\omega_0\alpha}{u}}-E_1(s)-e^{-s}\right)
 \nonumber \\
 & & -\frac{uG(0)}{2\omega_0\alpha} y(0)\left[2E_1\left(\frac{\omega_0\alpha}{u}\right)+
 2 G\left(\frac{\omega_0\alpha}{u}\right) +e^{-\frac{\omega_0\alpha}{u}}
 \ln\left(\frac{\omega_0\alpha}{u}\right)\right. \nonumber \\
 & & - \left. E_1\left(\frac{\omega_0\alpha}{u}\right)
 \ln \frac{\omega_0\alpha}{u} E_1(s)\ln s -2E_1(s)-2G(s) -e^{-s}\ln
 s\right]\label{eq:Z-perturb-y0}
\end{eqnarray}
\end{widetext}
Using the expansions of $E_1$ and $G$ for small argument:
\begin{eqnarray}
 E_1(s)=-\gamma-\ln s + o(1) \\
 G(s)=-\frac 1 2 (\ln s)^2 + \frac{\Gamma''(1)}{2} + o(1) \\
\end{eqnarray}
We find the following expression for $Y(l)$ at small
$uG(0)/(\omega_0\alpha)$:
\begin{widetext}
\begin{eqnarray}
 Y(l) &=& 1+\frac{u G(0)}{2\omega_0\alpha} \left[1-\gamma + \ln
 \frac{u}{\omega_0\alpha}\right]  -\epsilon y(0) \left[
 \ln^2 \left(\frac{u}{\omega_0\alpha}\right) + 2 (1-\gamma) \ln
 \left(\frac{u}{\omega_0\alpha} \right) \right. \nonumber \\ &&
 \left.+ \Gamma''(1)
 -2\gamma\right] -\left[ \frac{u G(0)}{2\omega_0\alpha}  + y(0) -\frac{u
    G(0)}{2\omega_0\alpha} y(0) \left[
 1-\gamma + \ln \left(\frac{ u}{\alpha
 \omega_0}\right)\right]\right] l\nonumber
\end{eqnarray}
\end{widetext}
For $y(0)<0$ a solution exists if:
\begin{eqnarray}
 \frac{u G(0)}{2\omega_0\alpha} > \frac{|y(0)|}{1+|y(0)| \ln \left(\frac{u
e^{1-\gamma}}{\alpha\omega_0}\right)}
\end{eqnarray}
In other words, the existence of the gap is controlled by the
ratio of the coupling constant to the marginally irrelevant
perturbation measured at the energy scale $\omega_0
e^{1-\gamma}$. We have\cite{abramowitz_math_functions}
$\Gamma''(1)=\gamma^2+\frac{\pi^2}{6}$. However, for what follows, it
is convenient to make an approximation $\Gamma''(1)\sim\gamma^2+1$.
Then, one can rewrite:
\begin{widetext}
\begin{eqnarray}
  \label{eq:Y-approximation}
  Y(l)=1+ \frac{u G(0)}{2\omega_0\alpha}\left(1-y(0)\ln\left(\frac{u
          e^{1-\gamma}}{\omega_0\alpha}\right)\right)\ln\left(\frac{u
          e^{1-\gamma}}{\omega_0\alpha}\right) -\left[y(0) + \frac{u G(0)}{2\omega_0\alpha}
    \left(1-y(0)\ln\left(\frac{u
          e^{1-\gamma}}{\omega_0\alpha}\right)\right)\right] l
\end{eqnarray}
\end{widetext}
The strong coupling scale is obtained for
$Y(l^*)=0$ which gives us:
\begin{equation}
\label{eq:scale-marginal-gap}
l^* = 1-\gamma+\ln \left(\frac u {\omega_0\alpha}\right) + \frac 1
{\frac{uG(0)}{2\omega_0\alpha} + \frac{y(0)}{1-y(0)\ln \left(\frac{u
        e^{1-\gamma}}{\omega_0\alpha}\right)}}
\end{equation}
We note that making $y(0)=0$ in the above formula gives back the
expression (\ref{eq:gap-antiad}).

Turning to the crossover to the adiabatic regime, we note that
again, due to the lack of a precise criterion for cutting the RG
flow, we can only propose an ansatz to relate the two regimes.
Extending the reasoning made in the previous discussion of the case
$y(0)=0$, we expect that the crossover happens when
$l_{ad.}^*-l_{antiad.}^*$ is minimal. The lengthscale $l^*_{ad.}$
has been obtained in section \ref{app:logar-corr-scal}. It reads:
\begin{equation}
  l^*_{ad.}=-\ln G(0) +\frac 3 2 \ln [1-|y(0)| \ln G(0)]
\end{equation}
Minimizing the difference of lengthscales then gives:
\begin{equation}\label{crossover-log}
  \frac{\frac{uG(0)}{2\omega_0\alpha}}{\left(\frac{uG(0)}{2\omega_0\alpha} -\frac{|y(0)|}{1+|y(0)|\ln \frac{u
        e^{1-\gamma}}{\omega_0\alpha}}\right)^2}=1+\frac 3 2
\frac{|y(0)|}{1-|y(0)|\ln G(0)}
\end{equation}
In the case of $y(0)\ll 1$, one can write $\frac u \alpha
G(0)=2\omega_0 (1+\epsilon)$. The quantity $\epsilon$ is
straightforward to obtain by expanding (\ref{crossover-log}). Finally,
one has:
\begin{multline}
  \frac u \alpha G(0)= \\
  2\omega_0\left[ 1 +2 \frac{|y(0)|}{1+|y(0)|\ln \frac{u
        e^{1-\gamma}}{\omega_0\alpha}} -\frac 3 2
    \frac{|y(0)|}{1+|y(0)|\ln \frac{u}{2\omega_0\alpha}}\right]
\end{multline}
Therefore, the crossover scale is only weakly affected by the presence
of logarithmic corrections.
An ansatz similar to Eq.~(\ref{eq:ansatz-gap-nologs}) can be derived
in the case of a spin chain with marginally irrelevant operator. Using
the results of Ref.~\cite{orignac04_spingap}, combined with the
App.~\ref{app:logar-corr-scal}, we obtain the following expression of
the gap in the adiabatic regime.
\begin{eqnarray} \label{eq:gap-quantitative-marginal-adia}
  \Delta = 1.96886\; \frac u \alpha G(0) \frac 1 {\left[1+|y(0)| \ln
  \frac{|y(0)|^{2/3}}{2.1557 G(0)}\right]^{\frac 3 2}},
\end{eqnarray}
Our calculation of the crossover scale shows that this expression is
valid provided $2\omega_0<uG(0)/\alpha$. For $2\omega_0>uG(0)/\alpha$,
an expression of the gap of the form $\Delta\sim {\cal C} e^{-l^*}$
with $l^*$ given by Eq.~(\ref{eq:scale-marginal-gap}) is
valid. Matching the two expressions for $2\omega_0=uG(0)/\alpha$
yields:
\begin{eqnarray} \label{eq:gap-quantitative-marginal-antiad}
  \Delta&=& 10.704\; \omega_0
  \frac{e^{\frac{|y(0)|}{1+|y(0)|\ln\left(\frac {ue^{-\gamma}}{\omega_0 \alpha}\right)}}}{\left[1+|y(0)| \ln
  \frac{|y(0)|^{2/3}u}{2 \times 2.1557\omega_0\alpha }\right]^{\frac 3
  2}}\nonumber \\ && \times \exp\left[\frac {-1}{\frac {uG(0)}{2\omega_0\alpha}
  -\frac{|y(0)|}{1+|y(0)|\ln\left(\frac {ue^{-\gamma}}{\omega_0
  \alpha}\right)}}\right]
\end{eqnarray}
for $2\omega_0><uG(0)/\alpha$. Letting $y(0)\to 0$, in
Eqs.~(\ref{eq:gap-quantitative-marginal-adia})-~(\ref{eq:gap-quantitative-marginal-antiad}),
we recover the previous formulas (\ref{eq:ansatz-gap-nologs}).


\begin{thebibliography}{10}
\bibitem{note_einstein}
Note that we are neglecting the interaction of the phonons with the $q\to 0$
  component of the energy density. This will lead us to predict a spin gap at
  high frequency $\omega_0$. However, strictly speaking, taking the Hamiltonian
  (\ref{hshp})-~(\ref{hspferm}), and performing the canonical transformation
  for $\omega_0\gg J$, no frustration is actually generated due to the local
  character of the Einstein phonons. This indicates that the current-current
  interaction terms induced by the $q\to0$ component may compensate those
  induced by the $2k_F$ term for sufficiently high frequency. However, since
  this effect does not exist for dispersive phonons, we will neglect the $q\to
  0$ contribution in what follows.
\end{thebibliography}

\end{document}